\documentclass[aps,reprint,superscriptaddress,amsmath,amssymb, floatfix,noeprint]{revtex4-2}

\usepackage{graphicx}
\usepackage[utf8]{inputenc}
\usepackage{siunitx}
\usepackage{physics}
\usepackage{tabularx}

\let\qty\SI

\newcommand{\calciumforty}{$^{40}\mathrm{Ca}^{+}$}
\newcommand{\strontium}{$^{88}\mathrm{Sr}^{+}$}
\newcommand{\bracket}[1]{\left|#1\right>}
\newcommand{\m}[1]{\mathrm{#1}}
\newcommand{\ketbraind}[3]{\ket{#1}_#2\! \bra{#3}}

\begin{document}
\bibliographystyle{apsrev4-1}
\title{Experimental realization of nonunitary multi-qubit operations}
\author{M. W. van Mourik}
\affiliation{Institut f\"ur Experimentalphysik, Universit\"at Innsbruck, Technikerstraße 25/4, 6020 Innsbruck, Austria}
\author{E. Zapusek}
\affiliation{Institute for Quantum Electronics, ETH Z\"urich, Otto-Stern-Weg 1, 8093 Z\"urich, Switzerland}
\author{P. Hrmo}
\author{L. Gerster}
\author{R. Blatt}
\affiliation{Institut f\"ur Experimentalphysik, Universit\"at Innsbruck, Technikerstraße 25/4, 6020 Innsbruck, Austria}
\author{T. Monz}
\affiliation{Institut f\"ur Experimentalphysik, Universit\"at Innsbruck, Technikerstraße 25/4, 6020 Innsbruck, Austria}
\affiliation{AQT, Technikerstraße 17, 6020 Innsbruck, Austria}
\author{P. Schindler}
\affiliation{Institut f\"ur Experimentalphysik, Universit\"at Innsbruck, Technikerstraße 25/4, 6020 Innsbruck, Austria}
\author{F. Reiter}
\affiliation{Institute for Quantum Electronics, ETH Z\"urich, Otto-Stern-Weg 1, 8093 Z\"urich, Switzerland}

\begin{abstract}
We demonstrate a novel experimental toolset that enables irreversible multi-qubit operations on a quantum platform. To exemplify our approach, we realize two elementary nonunitary operations: the OR and NOR gates. The electronic states of two trapped \calciumforty ions encode the logical information, and a co-trapped \strontium ion provides the irreversibility of the gate by a dissipation channel through sideband cooling. We measure 87\% and 81\% success rates for the OR and NOR gates, respectively. The presented methods are a stepping stone towards other nonunitary operations such as in quantum error correction and quantum machine learning.

%


\end{abstract}

\maketitle

\begin{figure}
    \includegraphics[width=\linewidth]{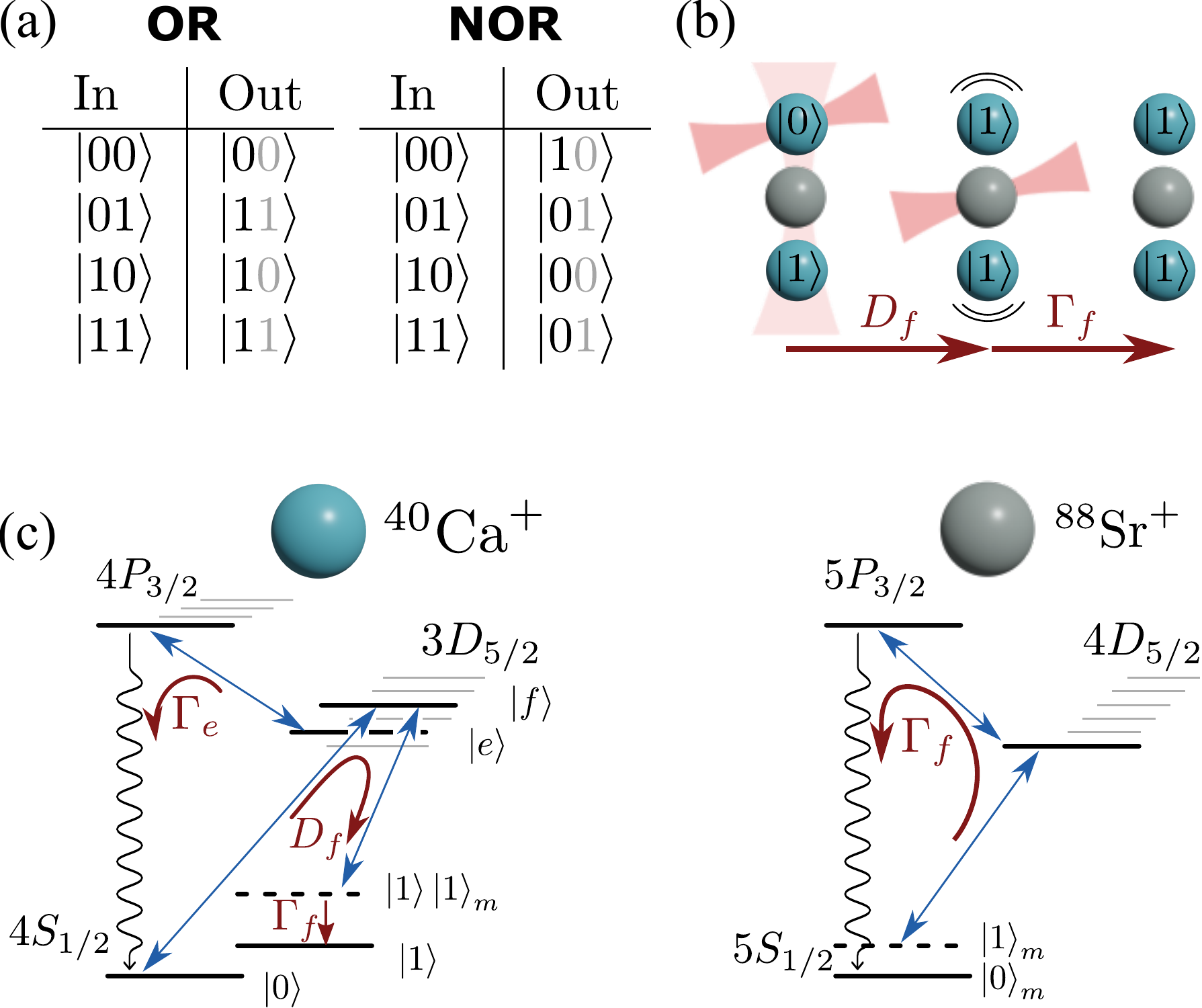}
    \caption{
    (a) Truth tables of the classical OR and NOR gates, with two-qubit output. The logical output is mapped on the left qubit. (b) Schematic representation of the OR gate acting on $\bracket{01}$: an engineered resonance process $D_f$ is a combination of a global and single-ion laser pulse, which together allow transfer to the desired state, $\bracket{11}$, plus an increase in motional mode occupation. This action is made irreversible by dissipation $\Gamma_f$ of this additional motion, by cooling a spectator ion species. (c) Overview of the relevant states in the data ions, \calciumforty, and cooling ion, \strontium. Logical bits $\bracket{0}$ and $\bracket{1}$ are stored in Calcium's $4S_{1/2}$ ground states. Auxiliary levels in the $3D_{5/2}$ manifold are used for engineered resonance transfer, $D_f$. Dissipation $\Gamma_e$ and $\Gamma_f$ occurs through spontaneous decay from $P_{3/2}$ to $S_{1/2}$, in Calcium and Strontium.
    }
    \label{fig:level_diagram}
\end{figure}

\label{sec:Introduction}
\textit{Introduction.---}
Classical computing is an immensely successful information processing paradigm. The success of computing can largely be explained by the rapid increase in computational power enabled by the miniaturization of the underlying circuits built from classical, irreversible gate operations (cf. Fig. \ref{fig:level_diagram}(a)). Today, the exponential growth of gate count on classical processors is reaching fundamental physical limits \cite{noauthor_international_2015}. In the continued pursuit of increasing computational power, a multitude of alternate technologies is being explored \cite{shulaker_carbon_2013,ma_small-diameter_2003,han_energy_2007,sugahara_spin_2004,salahuddin_use_2008,chen_integrated_2008,zhou_field_1997,qian_quantum_2014,lee_conceptual_2008,andreakou_optically_2014,feng_room_2005,currivan-incorvia_logic_2016}.
 
As an approach orthogonal to classical information processing quantum computing has recently received considerable attention.
Here, substantial advancements have been made, allowing for first demonstrations of essential ingredients such as quantum error correction
\cite{pino_demonstration_2021,kranzl_controlling_2022,egan_fault-tolerant_2021,postler_demonstration_2022,herrmann_realizing_2022,google_quantum_ai_exponential_2021}. This can be attributed to novel and advanced proposals and the continued improvement of established techniques \cite{olsacher_scalable_2020,zhang_submicrosecond_2020,krinner_realizing_2022,hrmo_native_2022,chu_scalable_2023}. Such advancements in controllability bring quantum computation closer to the ideal of an entirely unitary evolution towards the output state.
In certain algorithms, however, \emph{nonunitary} operations are required in combination with unitary quantum gates. 
Among these are algorithms for quantum machine learning, quantum optimization, and simulation, which are regarded as some of the most promising near-term applications for quantum information processing \cite{preskill_quantum_2018,bharti_noisy_2022,foldager_noise-assisted_2022,wu_variational_2019,zhu_generation_2020,wang_variational_2021,mazzola_nonunitary_2019,verdon_quantum_2019,wang_quantum_2011, del_re_driven-dissipative_2020, hu_quantum_2020,hu_general_2022,leppakangas_quantum_2022,cong_quantum_2019,herrmann_realizing_2022}.
Specifically, nonunitary operations are needed for the generation of low-temperature thermal states \cite{foldager_noise-assisted_2022,wu_variational_2019,zhu_generation_2020,wang_variational_2021}, as a projective filter \cite{mazzola_nonunitary_2019}, for the simulation of open systems \cite{verdon_quantum_2019,wang_quantum_2011, del_re_driven-dissipative_2020, hu_quantum_2020,hu_general_2022,leppakangas_quantum_2022}, or in quantum neural networks \cite{cong_quantum_2019,herrmann_realizing_2022}.
It has been suggested to implement nonunitary components through auxiliary qubits, randomized circuits, or mixed input states \cite{mazzola_nonunitary_2019,wang_variational_2021,wang_quantum_2011,del_re_driven-dissipative_2020,hu_general_2022,hu_quantum_2020,foldager_noise-assisted_2022, verdon_quantum_2019}.

Dissipation is inherently nonunitary, making it a natural choice for the creation of irreversible operations. The field of dissipation engineering, or reservoir engineering, uses the interaction of a quantum system with environmental degrees of freedom to achieve quantum information processing tasks \cite{harrington_engineered_2022, kronwald_dissipative_2014, agarwal_strong_2016, reiter_scalable_2016, herrmann_realizing_2022, kastoryano_dissipative_2011, lin_dissipative_2013, morigi_dissipative_2015, cole_dissipative_2021, cole_resource-efficient_2022, malinowski_generation_2021, doucet_high_2020, barreiro_open-system_2011, reiter_autonomous_2017, wang_autonomous_2019}.
Applications include state preparation by optical pumping, squeezing  \cite{kronwald_dissipative_2014,agarwal_strong_2016}, entanglement generation \cite{reiter_scalable_2016, kastoryano_dissipative_2011,  lin_dissipative_2013, morigi_dissipative_2015,  cole_dissipative_2021, cole_resource-efficient_2022, malinowski_generation_2021, doucet_high_2020}, quantum simulation \cite{barreiro_open-system_2011}, and quantum error correction \cite{reiter_autonomous_2017}. Dissipation towards the environment lifts the requirement for classical measurement and feedback, and it holds scaling and robustness advantages over unitary approaches \cite{kastoryano_dissipative_2011, lin_dissipative_2013, morigi_dissipative_2015}.
It has been formally shown that dissipation can be used to perform universal quantum computation \cite{verstraete_quantum_2009}. Still, so far dissipation engineering has mostly been focused on quantum state preparation and subspace stabilization.
We expand the possible set of applications taking a step towards a paradigm of general nonunitary quantum operations, by demonstrating the realization of irreversible classical gates (cf. Fig. \ref{fig:level_diagram}(a)) by means of engineered dissipation.

Here we present a physical realization of nonunitary operations in a trapped-ion system by use of dissipation engineering. 
By utilizing techniques from dissipation engineering, one can create nonunitary quantum gates that operate deterministically and without the need for ancilla qubits \cite{zapusek_nonunitary_2023}.
To this end, from quantum-mechanical interactions, we engineer the desired projective dynamics effecting classical gate operations. We implement a classical OR and NOR gate, whose truth tables are shown in Fig. \ref{fig:level_diagram}(a), where the output of the gate action is mapped onto the left qubit.
We employ selective coherent couplings to conditionally excite electronic states, utilizing the ions' shared motional modes, schematically outlined in Fig. \ref{fig:level_diagram}(b). Both sympathetic cooling and decay via an auxiliary level serve as the nonunitary components and complete the gate action. Experimentally, the desired dynamics can be implemented in a mixed-species trapped-ion system with single-qubit addressing capabilities \cite{brandl_cryogenic_2016}.
Through our work we show that carefully engineered nonunitary quantum dynamics have the potential to enrich the quantum engineer's toolbox, by performing a broad class of operations.

\label{sec:operation}
\textit{Principle of operation.---}
Two co-trapped \calciumforty ions serve as information carriers, with the logical states $\bracket{0}$ and $\bracket{1}$ encoded in the $4S_{1/2}(m=-1/2)$ and $(m=+1/2)$ Zeeman sub-levels (see Fig. \ref{fig:level_diagram}(c)). The ions are trapped in a harmonic potential, and share motional modes. We use the notation $\bracket{i}_1 \otimes \bracket{j}_2\otimes \bracket{n}_m= \bracket{ij}\bracket{n}_m$, for electronic states $i$ and $j$ of ions 1 and 2, and mode occupation $n$ of a specific motional mode $m$. For brevity, the mode occupation $n$ is often omitted in our notation when $n=0$, i.e. $\bracket{ij}=\bracket{ij}\bracket{0}_m$. As seen in the truth-table in Fig. \ref{fig:level_diagram}(a), the OR gate corresponds to the mapping of $\ket{01}$ to $\ket{11}$, which is analogous to the condition that the first qubit is flipped from $\bracket{0}$ to $\bracket{1}$ if and only if the second qubit is the state $\ket{1}$. The desired conditional operation is augmented by making use of a specific motional mode to encode information about the parity of the system. Access to motional modes is enabled through the auxiliary state $\bracket{f}$, for which we use the metastable $3D_{5/2}(m=+1/2)$ level. The $4S_{1/2}\leftrightarrow 3D_{5/2}$ transition to this auxiliary state is coupled with coherent 729 nm light.

\begin{figure}
    \includegraphics[width=\linewidth]{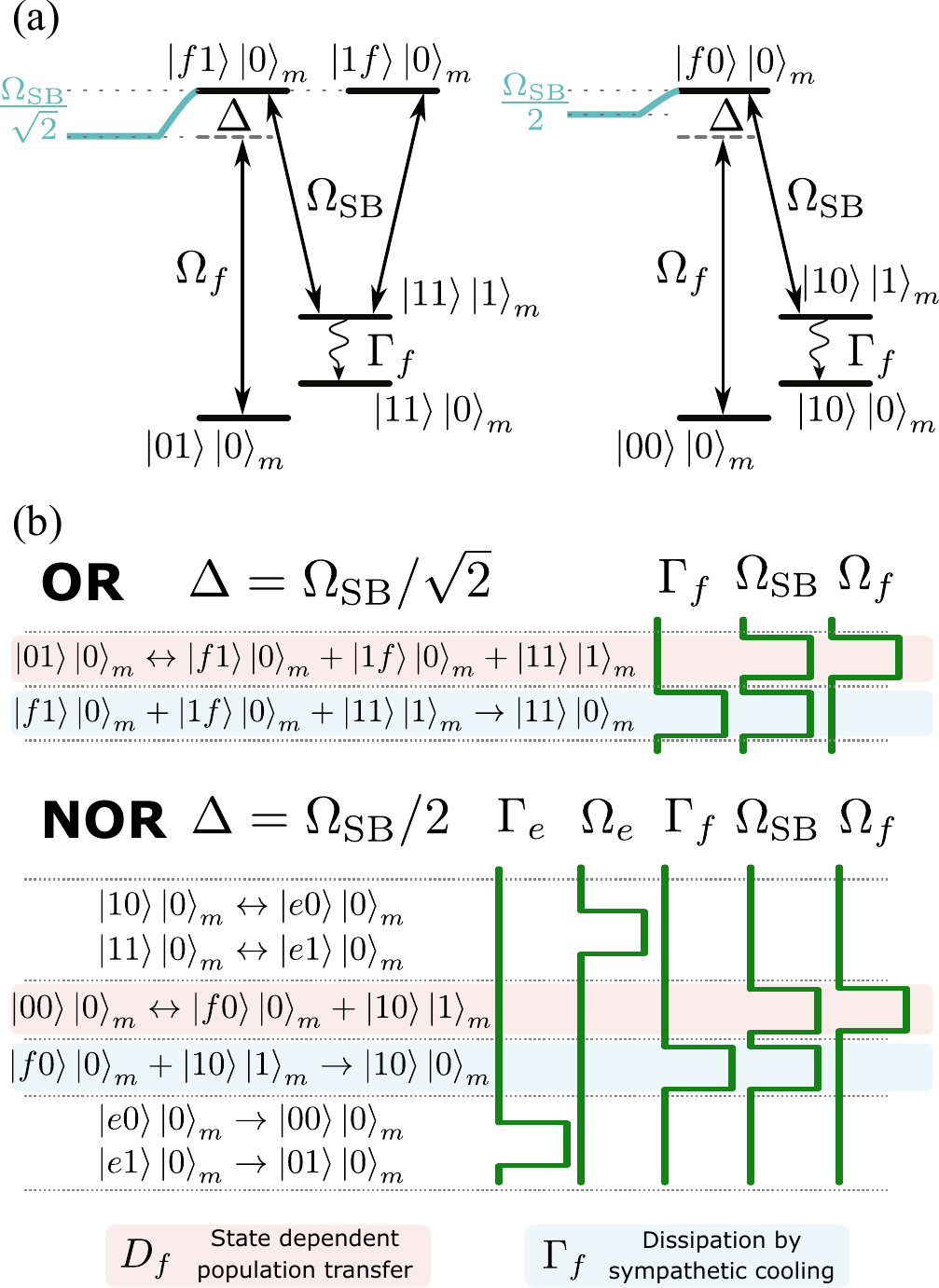}
    \caption{Overview of the gate mechanism. (a) Desired (left) and undesired (right) process for the OR gate. State $\ket{01}\ket{0}_m$ is off-resonantly driven to $\ket{f1}\ket{0}_m$ with Rabi frequency $\Omega_f$. Due to the coupling $\Omega_{\m{SB}}$, $\ket{f1}\ket{0}_m$ is  hybridized with states $\ket{11}\ket{1}_m$ and $\ket{1f}\ket{0}$, resulting in a dressed state splitting. For $\Delta=\Omega_{\m{SB}}/\sqrt{2}$, the carrier drive is on resonance with the dressed state, and is therefore excited. The gate action is completed by sympathetically cooling the motional mode. The dressed states $\ket{f0}\ket{0}_m$ and $\ket{10}\ket{1}_m$, accessible from $\ket{00}\ket{0}_m$, are shifted by $\pm \Omega_{\m{SB}}/2$, and are therefore not in resonance with the carrier drive. Excitation from $\ket{00}\ket{0}_m$ is thus suppressed. (b) Pulse sequence of OR and NOR gates. Transitions used in the experiment are indicated in Fig. \ref{fig:level_diagram}(c). $\Omega_f$ and $\Omega_e$ act on the first ion, coupling $\bracket{0}$ with $\bracket{f}$ and $\bracket{1}$ with $\bracket{e}$. $\Omega_{\m{SB}}$ acts on both ions, coupling the red sideband of $\bracket{1}$ and $\bracket{f}$.}
    \label{fig:sequence_schematic}
\end{figure}

The population transfer mechanism is summarized below (details in Supplemental material \footnote{The equations describing the dynamics of the system are given in the Supplemental material, Section A}). A drive with Rabi frequency $\Omega_f$ and detuning $\Delta$ is applied to the first ion's $\bracket{0}\leftrightarrow\bracket{f}$ transition. We refer to this drive as the probe. Without any further couplings, the probe would excite the two states $\bracket{00}$ and $\bracket{01}$, and leave the states $\bracket{10}$ and $\bracket{11}$ unchanged. A second drive, which we refer to as the sideband drive, is applied to both ions on the $\bracket{f}\leftrightarrow\bracket{1}$ transition, though red-detuned by the frequency of the motional mode. This drive therefore couples the states $\ket{f}\ket{n}_m\leftrightarrow\ket{1}\ket{n+1}_m$. In particular, the transition between $\bracket{f}\ket{0}_m \leftrightarrow \ket{1}\ket{1}_m$ occurs with Rabi frequency $\Omega_{\m{SB}}$. Under the condition that $\Omega_{\m{SB}}\gg\Omega_{f}$, the states excited from $\ket{01}\ket{0}_m$ form dressed states $(\bracket{f1}\ket{0}_m + \ket{1f}\ket{0}_m \pm \sqrt{2}\bracket{11}\ket{1}_m)/2$.
These dressed states have a frequency shift of $\pm\Omega_{\m{SB}}/\sqrt{2}$, with respect to the bare $\ket{f1}\ket{0}_m$ state, as shown in Fig. \ref{fig:sequence_schematic}(a), left.
In contrast, the initial state $\ket{00}\ket{0}_m$ is excited to the dressed states $(\ket{f0}\ket{0}_m \pm \ket{10}\ket{1}_m)/\sqrt{2}$ (Fig. \ref{fig:sequence_schematic}(a), right), which reside at frequencies $\pm \Omega_{\m{SB}}/2$.
The three-level dressed states have increased frequency shifts compared to the two-level dressed states, because of constructive interference of the couplings $\{\ket{1f}\ket{0}_m, \ket{f1}\ket{0}_m\}  \leftrightarrow \ket{11}\ket{1}_m$.
Choosing a probe pulse detuning $\Delta=\Omega_{\m{SB}}/\sqrt{2}$ therefore enables excitation from $\bracket{01}$, while excitation from $\bracket{00}$ is out of resonance and thus suppressed. This population transfer through engineered resonance is denoted by $D_f$ in Fig. \ref{fig:level_diagram}(b) and (c).

The conditional excitation is made nonunitary by a decay process enabled by sideband cooling a co-trapped \strontium ion, indicated in Fig. \ref{fig:level_diagram}(c). In the OR gate, population that cycles through $\ket{11}\ket{1}_m$ is dissipatively transferred to $\ket{11}\ket{0}_m$ at a rate $\Gamma_f$. Since $\ket{11}\ket{0}_m$ does not couple with either the sideband drive nor the probe, population remains in this state, thus completing the transfer $\ket{01}\rightarrow\ket{11}$. In order to avoid interfering with the excitation during the probe process, the dissipation is realized in a subsequent step \cite{lin_dissipative_2013,malinowski_generation_2021}. 

We expand this mechanism to the universal NOR gate, whose truth table is shown in Fig.\,\ref{fig:level_diagram}(a). This gate can be constructed by concatenating a unitary NOT gate with the OR gate. However, we present a fully dissipative implementation where the mapping $\ket{00}\rightarrow\ket{10}$ follows the same procedure as the OR gate. In contrast to the OR gate, for the NOR gate we use the detuning $\Delta=\Omega_{\m{SB}}/2$ to excite the initial state $\ket{00}$. In addition, the transfers $\ket{11}\rightarrow\ket{01}$ and $\ket{10}\rightarrow\ket{00}$ are required for the gate action which can be implemented by a single-qubit dissipative process. Both mappings are achieved by optically pumping the first ion from $\ket{1}$ to $\ket{0}$ over another auxiliary level, $3D_{5/2}(m=-3/2)\equiv \ket{e}$, and subsequently to $4P_{3/2}(m=-3/2)$, from where spontaneous decay returns population to the $\bracket{0}$ state.

\label{sec:Experimental_overview}
\textit{Experimental overview.---}
The experiments have been carried out with a segmented surface trap in a cryogenic environment \cite{brandl_cryogenic_2016}. Ions are stored in the Ca-Sr-Ca configuration. Collisions with particles in the background gas may disrupt this orientation. Therefore, we periodically apply a sequence of voltages to the trap electrodes that deterministically place the ions back in the desired configuration \cite{jost_entangled_2009,home_quantum_2013}.

We set the sideband drive to couple to the crystal's axial in-phase (ip) mode. 
Confining potentials are set so that the in-phase mode frequency is $\omega_{\mathrm{ip}}/(2\pi)=\qty{550}{\kilo\hertz}$. This value is chosen as a trade-off between ensuring a low motional mode heating rate ($\dot{\bar{n}}\propto\omega^{-\alpha}_{\mathrm{ip}}$ with $\alpha \approx 2$) and a sufficiently high coupling to the motional mode $\Omega_{\m{SB}}$ through laser interaction ($\Omega_{\m{SB}}\propto\omega^{-1/2}_{\mathrm{ip}}$) \cite{Brownnutt2015}. At this frequency, we have measured an axial in-phase mode heating rate of 106(20) phonons per second, and an initial mean mode occupation of 0.14 phonons after sideband cooling. Both the axial in- and out-of-phase modes of the ion crystal are sideband cooled.  
Ions are initialized in $\bracket{00}$ using optical pumping. We prepare the remaining possible initial states $\bracket{01}$, $\bracket{10}$, and $\bracket{11}$, using a combination of single-ion and collective $\pi$-pulses on the $\bracket{0}\leftrightarrow\bracket{f}$ and $\bracket{f}\leftrightarrow\bracket{1}$ transitions.

The sequences of operations for the OR and NOR operations are schematically shown in Fig. \ref{fig:sequence_schematic}(b), referring to the states shown in Fig. \ref{fig:level_diagram}(c), with $\Omega_f$ and $\Omega_e$ acting only on the first ion, coupling $\bracket{0}$ with $\bracket{f}$ and $\bracket{1}$ with $\bracket{e}$, respectively. $\Omega_{\m{SB}}$ acts on both ions, and couples the red sideband of $\bracket{1}$ and $\bracket{f}$.

For the OR operation, the initial state $\bracket{01}$ is to be transferred to $\bracket{11}$, while all other initial states remain unchanged.
The dressed state splitting is produced with a sideband drive with Rabi frequency $\Omega_{\m{SB}}/(2\pi)\approx\qty{8}{\kilo\hertz}$. The probe beam, simultaneously applied to only the first ion, is detuned by $\Delta=\Omega_{\m{SB}}/\sqrt{2}$ from the $\bracket{0}\leftrightarrow\bracket{f}$ carrier transition, with an on-resonance Rabi frequency $\Omega_f/(2\pi)\approx\qty{1.15}{\kilo\hertz}$. These pulses are applied for a duration of $2\pi/\Omega_f=\qty{900}{\micro\second}$, which excites $\bracket{01}$ to the dressed state as shown in Fig. \ref{fig:sequence_schematic}(a) left. Similar transfer from an initial state of $\bracket{00}$ is suppressed because the resonance condition, shown in Fig. \ref{fig:sequence_schematic}(a) right, is not met.

Following this state-dependent population transfer, the state $\ket{11}\ket{1}_m$ is dissipatively transferred to $\ket{11}\ket{0}_m$ by cooling the Sr ion.
The sideband coupling $\Omega_{\m{SB}}$ is maintained during the cooling step, which fully depletes the populated dressed state.

The NOR gate follows a similar procedure as above, shown in Fig. \ref{fig:sequence_schematic}(b), though now population transfer from $\bracket{00}$ is enabled by choosing $\Delta=\Omega_{\mathrm{SB}}/2$, which suppresses excitation from $\bracket{01}$. The additional channel of dissipation required by the NOR operation, $\bracket{1}\rightarrow\bracket{0}$ for only the first ion, is performed in multiple steps since it would otherwise conflict with the simultaneously required $\bracket{00}\rightarrow\bracket{10}$ operation. Preceding the engineered dissipation, population in the first ion's $\bracket{1}$ state is stored in $\bracket{e}$. After the engineered dissipation, $\sigma^-$-polarized light at 854 nm transfers population in $\bracket{e}$ to the $4P_{3/2}$ level, favoring the $m=-3/2$ Zeeman sublevel, from which spontaneous decay brings it to $\bracket{0}$.

At the end of the sequence, the population is read out with state-dependent fluorescence detection using an EMCCD camera, which distinguishes excitation of the $S$ and $D$ manifolds. As the logical information is carried in the two $S$-levels (cf. Fig.\ref{fig:level_diagram}(c)), the population in $\bracket{1}$  needs to be transferred to $\bracket{e}$ before the measurement. This state-readout does not differentiate between between $\bracket{1}$ and $\bracket{f}$. We use the notation $P_{ij}$ to indicate the population in state $\bracket{ij}$. We can separately measure the occupation of the motional mode by applying a pulse on resonance with either the red or blue sideband of one of the Strontium ion's $5S_{1/2}\leftrightarrow4D_{5/2}$ transitions, and reading out its state \cite{Brownnutt2015}. The difference of the excitation probability of the red and blue sideband excitations is used to infer the population in the motional ground state.

\begin{figure}
    \includegraphics[width=\linewidth]{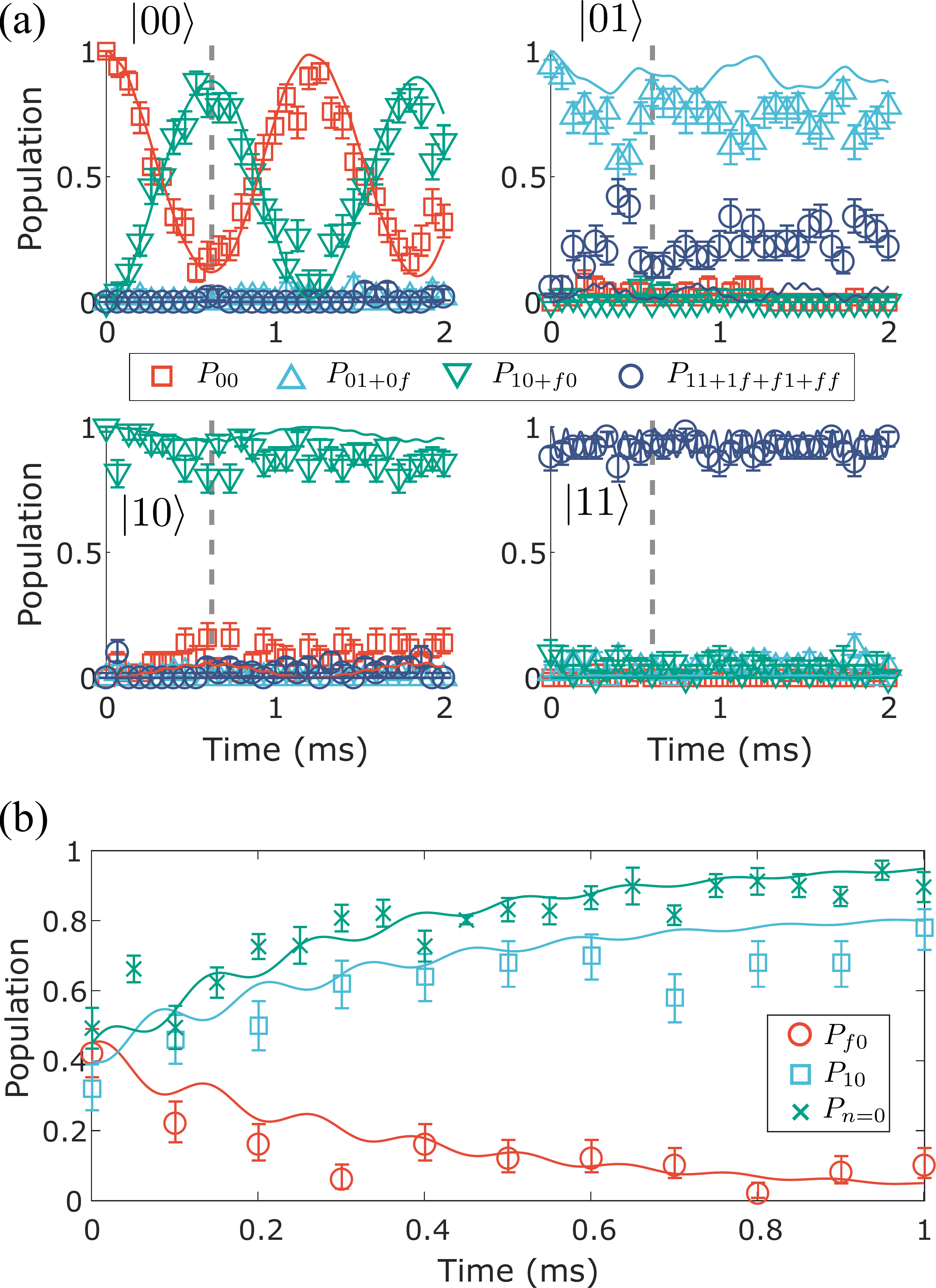}
    \caption{(a) Experimental demonstration of state-dependent population transfer transfer with $\Delta=\Omega_{\m{SB}}/2$, shown for each possible initial state. The lines indicate simulated results, which include measured initial phonon number and heating rate as simulation parameters. (b) Demonstration of dissipation, after a maximal probe transfer from $\bracket{00}$ (at a time marked by the gray dashed line in (a) ). We show the evolution of $P_{f0}$ and $P_{10}$, and use sideband thermometry on the Strontium ion to infer the ground-state phonon occupation. The lines show simulated results, in which the dissipation rate $\Gamma_f$ is obtained through a least-squares fit between simulated and measured data.}
    \label{fig:results_pulses}
\end{figure}

\label{sec:Results}
\textit{Results.---}
We first demonstrate the central building block of resonance engineering, the state-dependent population transfer, by showing its time-evolution, using a detuning of $\Delta=\Omega_{\m{SB}}/2$.
The change in population is shown for all four initial states. The intended behavior, Rabi cycling from $\bracket{00}$ and no transfer from the other initial states, is apparent in Fig. \ref{fig:results_pulses}(a). The solid lines denote simulated data.
The simulations numerically solve the system's master equation  \cite{Note1}, and use experimentally determined parameters described above, including the initial phonon number and heating rate.
The gray dashed line marks the duration of the pulse with the maximum state transfer, \qty{600}{\micro\second}, where 82\% of population has depleted from $\bracket{00}$, and only 16\% from $\bracket{01}$. The deviation from a full population transfer is attributed to the non-zero initial phonon number and heating rate, corroborated by the simulated results.

After this population transfer, the state $(\ket{10}\ket{1}_m+\ket{f0}\ket{0}_m)/\sqrt{2}$ should be dissipatively transferred to $\ket{10}\ket{0}_m$.
We demonstrate this process by showing the evolution of the populations $P_{f0}$ and $P_{10}$ and the phonon ground state occupation over time in Fig. \ref{fig:results_pulses}(b). The electronic states are differentiated by running the measurement twice: once with the transfer of $\ket{1}$ to $\ket{f}$, and once without. The latter measurement does not discriminate between $\bracket{0}$ and $\bracket{1}$. The population $P_{10}$ is inferred from the difference between the first and second measurement. We additionally measure the phonon occupation. The lines are simulated results, using the same sideband coupling strength $\Omega_{\mathrm{SB}}$ as in (a). A dissipation rate of  $\Gamma_f=\qty{4.5(6)}{\kilo\hertz}$ is determined by a least-squares fit between the simulated and measured results. After \qty{1}{\milli\second} of applying the dissipation pulse, approximately 80\% of the population is in $\ket{10}\ket{0}_m$. 
The population is trapped there because of the irreversible nature of the dissipation.

\begin{figure}
    \includegraphics[width=\linewidth]{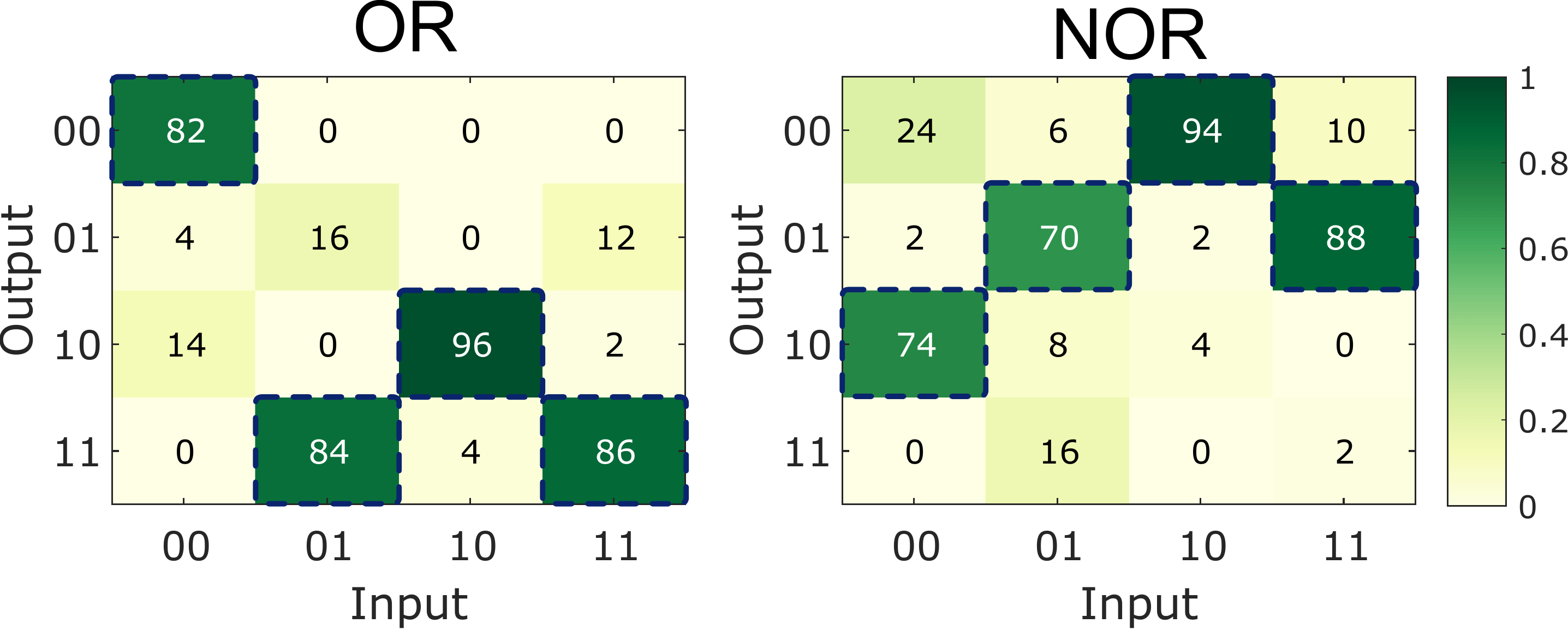}
    \caption{Measured population truth tables of the OR and NOR gates, with the intended output states marked with dashed lines. Values are in percent, and are determined from 50 experimental shots for each input setting. The OR and NOR gates have an average population fidelity of 87(5)\% and 81(5)\%.}
    \label{fig:results_truthtable}
\end{figure}

Having demonstrated and characterized the engineered resonance and dissipation processes, we apply these steps within the full pulse sequences shown in Fig \ref{fig:sequence_schematic}(b) to perform the OR and NOR gates. 
Figure \ref{fig:results_truthtable} shows the measured population outcome for each of the four possible initial states for the OR and NOR gates. Populations are determined from 50 experimental repetitions for each input state. Both truth tables exhibit the intended gate behavior: for all input states, the majority of the population is transferred to the desired state, marked in the figure with dashed boxes. 
The initial states $\bracket{01}$ and $\bracket{00}$ are transferred following the engineered resonance scheme for the OR and NOR gates, and have success rates of 84(5)\% and 74(6)\%. As confirmed by simulations and analytics \footnote{An analytic expression for the error is calculated in the Supplemental material, Section B}, the primary source of error is attributed to a non-zero initial phonon number and the heating rate. Since the coupling strength to a motional sideband is dependent on the phonon number, the resonance condition of the engineered population transfer is not met for $n\geq1$.

\label{sec:Outlook}
\textit{Conclusion and Outlook.---}
We have implemented nonunitary multi-qubit operations in a trapped-ion system by use of engineered dissipation. The schemes for the OR and NOR gate performed the operations with average fidelities of 87(5)\% and 81(5)\%, respectively. This constitutes the first realization of dissipative quantum gate operations. The leading source of error stems from heating and thus imperfect cooling of the mode over which the intended dissipation process occurs.

This heating process is the result of electronic noise on the trap surface, and by phonon transfer from other uncooled motional modes caused by mode-coupling, both of which are known challenges of microfabricated ion traps \cite{boldin_measuring_2018,home_normal_2011}, and is further exacerbated by complications involved in mixed-species operation \cite{home_quantum_2013}. Such issues are technical, and do not pose fundamental limitations: Future experiments could implement improved trap design and manufacturing to reduce heating due to technical noise, and improved cooling techniques such as polarization gradient cooling \cite{Joshi2020} and electromagnetically-induced-transparancy (EIT) cooling \cite{Morigi2000}. Much like recent dissipative high-fidelity schemes for entangled state preparation\,\cite{malinowski_generation_2021, cole_resource-efficient_2022, cole_dissipative_2021, doucet_high_2020} improved upon the fidelities of their first-generation counterparts we would expect future implementations to improve the fidelity.

Nonunitary operations are of relevance in a wide range of quantum information algorithms.
For example, in a NISQ context, quantum convolutional neural networks use measurements and conditional feedback operations to process information \cite{cong_quantum_2019,herrmann_realizing_2022}. These elements could be replaced by integrated nonunitary operations, thereby avoiding classical measurements and feedforward.

Regarding universal fault-tolerant quantum computation, quantum error correction also constitutes a nonunitary process as multiple erroneous processes are mapped to the same corrected state. 
Our work can be seen as a stepping stone towards an implementation of autonomous quantum error correction, in which erroneous states are coherently mapped to oscillator excitations and are then removed through dissipation \cite{reiter_autonomous_2017}. 
We have demonstrated the required techniques, resonance engineering and sympathetic cooling, in the present experiment.

\section*{Acknowledgments}
We gratefully acknowledge discussions with Jonathan Home and support by the EU Quantum Technology Flagship grant AQTION under Grant Agreement number 820495, and by the US Army Research Office through Grant No. W911NF-14-1-010 and W911NF-21-1-0007. 
We also acknowledge funding by the Austrian Science Fund (FWF), through the SFB BeyondC (FWF Project No. F7109), ERC-2020-STG 948893, and by the IQI GmbH. The research is also based upon work supported by the Office of the Director of National Intelligence (ODNI), Intelligence Advanced Research Projects Activity (IARPA), via the US Army Research Office Grant No. W911NF-16-1-0070.
F.R. and E.Z. acknowledge funding from the Swiss National Science Foundation (Ambizione grant no. PZ00P2 186040) and the ETH Research Grant ETH-49 20-2. 
\paragraph*{\textbf{Author Contributions}}
E.Z. developed, under the guidance of F.R., the theoretical protocol. M.v.M, F.R., P.S., and E.Z. designed the experiment. M.v.M. carried out the experiment and analyzed the data. M.v.M., P.H., and L.G. contributed to the experimental setup. R.B., T.M., F.R., P.S. supervised the project. All authors contributed to the manuscript.

\bibliography{main,BibMvM}

\begin{thebibliography}{64}%
\makeatletter
\providecommand \@ifxundefined [1]{%
 \@ifx{#1\undefined}
}%
\providecommand \@ifnum [1]{%
 \ifnum #1\expandafter \@firstoftwo
 \else \expandafter \@secondoftwo
 \fi
}%
\providecommand \@ifx [1]{%
 \ifx #1\expandafter \@firstoftwo
 \else \expandafter \@secondoftwo
 \fi
}%
\providecommand \natexlab [1]{#1}%
\providecommand \enquote  [1]{``#1''}%
\providecommand \bibnamefont  [1]{#1}%
\providecommand \bibfnamefont [1]{#1}%
\providecommand \citenamefont [1]{#1}%
\providecommand \href@noop [0]{\@secondoftwo}%
\providecommand \href [0]{\begingroup \@sanitize@url \@href}%
\providecommand \@href[1]{\@@startlink{#1}\@@href}%
\providecommand \@@href[1]{\endgroup#1\@@endlink}%
\providecommand \@sanitize@url [0]{\catcode `\\12\catcode `\$12\catcode
  `\&12\catcode `\#12\catcode `\^12\catcode `\_12\catcode `\%12\relax}%
\providecommand \@@startlink[1]{}%
\providecommand \@@endlink[0]{}%
\providecommand \url  [0]{\begingroup\@sanitize@url \@url }%
\providecommand \@url [1]{\endgroup\@href {#1}{\urlprefix }}%
\providecommand \urlprefix  [0]{URL }%
\providecommand \Eprint [0]{\href }%
\providecommand \doibase [0]{http://dx.doi.org/}%
\providecommand \selectlanguage [0]{\@gobble}%
\providecommand \bibinfo  [0]{\@secondoftwo}%
\providecommand \bibfield  [0]{\@secondoftwo}%
\providecommand \translation [1]{[#1]}%
\providecommand \BibitemOpen [0]{}%
\providecommand \bibitemStop [0]{}%
\providecommand \bibitemNoStop [0]{.\EOS\space}%
\providecommand \EOS [0]{\spacefactor3000\relax}%
\providecommand \BibitemShut  [1]{\csname bibitem#1\endcsname}%
\let\auto@bib@innerbib\@empty
\bibitem [{noa(2015)}]{noauthor_international_2015}%
  \BibitemOpen
  \href {www.semiconductors.org/wp-content/
  uploads/2018/06/0_2015-ITRS-2.0-Executive-Report-1.pdf} {\emph {\bibinfo
  {title} {International {{Technology Roadmap}} for {{Semiconductors}} 2.0}}},\
  \bibinfo {type} {Tech. Rep.}\ (\bibinfo  {institution} {{ITRS}},\ \bibinfo
  {year} {2015})\BibitemShut {NoStop}%
\bibitem [{\citenamefont {Shulaker}\ \emph {et~al.}(2013)\citenamefont
  {Shulaker}, \citenamefont {Hills}, \citenamefont {Patil}, \citenamefont
  {Wei}, \citenamefont {Chen}, \citenamefont {Wong},\ and\ \citenamefont
  {Mitra}}]{shulaker_carbon_2013}%
  \BibitemOpen
  \bibfield  {author} {\bibinfo {author} {\bibfnamefont {M.~M.}\ \bibnamefont
  {Shulaker}}, \bibinfo {author} {\bibfnamefont {G.}~\bibnamefont {Hills}},
  \bibinfo {author} {\bibfnamefont {N.}~\bibnamefont {Patil}}, \bibinfo
  {author} {\bibfnamefont {H.}~\bibnamefont {Wei}}, \bibinfo {author}
  {\bibfnamefont {H.-Y.}\ \bibnamefont {Chen}}, \bibinfo {author}
  {\bibfnamefont {H.-S.~P.}\ \bibnamefont {Wong}}, \ and\ \bibinfo {author}
  {\bibfnamefont {S.}~\bibnamefont {Mitra}},\ }\href {\doibase
  10.1038/nature12502} {\bibfield  {journal} {\bibinfo  {journal} {Nature}\
  }\textbf {\bibinfo {volume} {501}},\ \bibinfo {pages} {526} (\bibinfo {year}
  {2013})}\BibitemShut {NoStop}%
\bibitem [{\citenamefont {Ma}\ \emph {et~al.}(2003)\citenamefont {Ma},
  \citenamefont {Lee}, \citenamefont {Au}, \citenamefont {Tong},\ and\
  \citenamefont {Lee}}]{ma_small-diameter_2003}%
  \BibitemOpen
  \bibfield  {author} {\bibinfo {author} {\bibfnamefont {D.~D.~D.}\
  \bibnamefont {Ma}}, \bibinfo {author} {\bibfnamefont {C.~S.}\ \bibnamefont
  {Lee}}, \bibinfo {author} {\bibfnamefont {F.~C.~K.}\ \bibnamefont {Au}},
  \bibinfo {author} {\bibfnamefont {S.~Y.}\ \bibnamefont {Tong}}, \ and\
  \bibinfo {author} {\bibfnamefont {S.~T.}\ \bibnamefont {Lee}},\ }\href
  {\doibase 10.1126/science.1080313} {\bibfield  {journal} {\bibinfo  {journal}
  {Science}\ }\textbf {\bibinfo {volume} {299}},\ \bibinfo {pages} {1874}
  (\bibinfo {year} {2003})}\BibitemShut {NoStop}%
\bibitem [{\citenamefont {Han}\ \emph {et~al.}(2007)\citenamefont {Han},
  \citenamefont {{\"O}zyilmaz}, \citenamefont {Zhang},\ and\ \citenamefont
  {Kim}}]{han_energy_2007}%
  \BibitemOpen
  \bibfield  {author} {\bibinfo {author} {\bibfnamefont {M.~Y.}\ \bibnamefont
  {Han}}, \bibinfo {author} {\bibfnamefont {B.}~\bibnamefont {{\"O}zyilmaz}},
  \bibinfo {author} {\bibfnamefont {Y.}~\bibnamefont {Zhang}}, \ and\ \bibinfo
  {author} {\bibfnamefont {P.}~\bibnamefont {Kim}},\ }\href {\doibase
  10.1103/PhysRevLett.98.206805} {\bibfield  {journal} {\bibinfo  {journal}
  {Phys. Rev. Lett.}\ }\textbf {\bibinfo {volume} {98}},\ \bibinfo {pages}
  {206805} (\bibinfo {year} {2007})}\BibitemShut {NoStop}%
\bibitem [{\citenamefont {Sugahara}\ and\ \citenamefont
  {Tanaka}(2004)}]{sugahara_spin_2004}%
  \BibitemOpen
  \bibfield  {author} {\bibinfo {author} {\bibfnamefont {S.}~\bibnamefont
  {Sugahara}}\ and\ \bibinfo {author} {\bibfnamefont {M.}~\bibnamefont
  {Tanaka}},\ }\href {\doibase 10.1063/1.1689403} {\bibfield  {journal}
  {\bibinfo  {journal} {Appl. Phys. Lett.}\ }\textbf {\bibinfo {volume} {84}},\
  \bibinfo {pages} {2307} (\bibinfo {year} {2004})}\BibitemShut {NoStop}%
\bibitem [{\citenamefont {Salahuddin}\ and\ \citenamefont
  {Datta}(2008)}]{salahuddin_use_2008}%
  \BibitemOpen
  \bibfield  {author} {\bibinfo {author} {\bibfnamefont {S.}~\bibnamefont
  {Salahuddin}}\ and\ \bibinfo {author} {\bibfnamefont {S.}~\bibnamefont
  {Datta}},\ }\href {\doibase 10.1021/nl071804g} {\bibfield  {journal}
  {\bibinfo  {journal} {Nano Lett.}\ }\textbf {\bibinfo {volume} {8}},\
  \bibinfo {pages} {405} (\bibinfo {year} {2008})}\BibitemShut {NoStop}%
\bibitem [{\citenamefont {Chen}\ \emph {et~al.}(2008)\citenamefont {Chen},
  \citenamefont {Kam}, \citenamefont {Markovic}, \citenamefont {Liu},
  \citenamefont {Stojanovic},\ and\ \citenamefont
  {Alon}}]{chen_integrated_2008}%
  \BibitemOpen
  \bibfield  {author} {\bibinfo {author} {\bibfnamefont {F.}~\bibnamefont
  {Chen}}, \bibinfo {author} {\bibfnamefont {H.}~\bibnamefont {Kam}}, \bibinfo
  {author} {\bibfnamefont {D.}~\bibnamefont {Markovic}}, \bibinfo {author}
  {\bibfnamefont {T.-J.~K.}\ \bibnamefont {Liu}}, \bibinfo {author}
  {\bibfnamefont {V.}~\bibnamefont {Stojanovic}}, \ and\ \bibinfo {author}
  {\bibfnamefont {E.}~\bibnamefont {Alon}},\ }in\ \href {\doibase
  10.1109/ICCAD.2008.4681660} {\emph {\bibinfo {booktitle} {2008 {{IEEE}}/{{ACM
  International Conference}} on {{Computer-Aided Design}}}}}\ (\bibinfo
  {publisher} {{IEEE}},\ \bibinfo {address} {{San Jose, CA, USA}},\ \bibinfo
  {year} {2008})\ pp.\ \bibinfo {pages} {750--757}\BibitemShut {NoStop}%
\bibitem [{\citenamefont {Zhou}\ \emph {et~al.}(1997)\citenamefont {Zhou},
  \citenamefont {Newns}, \citenamefont {Misewich},\ and\ \citenamefont
  {Pattnaik}}]{zhou_field_1997}%
  \BibitemOpen
  \bibfield  {author} {\bibinfo {author} {\bibfnamefont {C.}~\bibnamefont
  {Zhou}}, \bibinfo {author} {\bibfnamefont {D.~M.}\ \bibnamefont {Newns}},
  \bibinfo {author} {\bibfnamefont {J.~A.}\ \bibnamefont {Misewich}}, \ and\
  \bibinfo {author} {\bibfnamefont {P.~C.}\ \bibnamefont {Pattnaik}},\ }\href
  {\doibase 10.1063/1.118285} {\bibfield  {journal} {\bibinfo  {journal} {Appl.
  Phys. Lett.}\ }\textbf {\bibinfo {volume} {70}},\ \bibinfo {pages} {598}
  (\bibinfo {year} {1997})}\BibitemShut {NoStop}%
\bibitem [{\citenamefont {Qian}\ \emph {et~al.}(2014)\citenamefont {Qian},
  \citenamefont {Liu}, \citenamefont {Fu},\ and\ \citenamefont
  {Li}}]{qian_quantum_2014}%
  \BibitemOpen
  \bibfield  {author} {\bibinfo {author} {\bibfnamefont {X.}~\bibnamefont
  {Qian}}, \bibinfo {author} {\bibfnamefont {J.}~\bibnamefont {Liu}}, \bibinfo
  {author} {\bibfnamefont {L.}~\bibnamefont {Fu}}, \ and\ \bibinfo {author}
  {\bibfnamefont {J.}~\bibnamefont {Li}},\ }\href {\doibase
  10.1126/science.1256815} {\bibfield  {journal} {\bibinfo  {journal}
  {Science}\ }\textbf {\bibinfo {volume} {346}},\ \bibinfo {pages} {1344}
  (\bibinfo {year} {2014})}\BibitemShut {NoStop}%
\bibitem [{\citenamefont {Lee}\ and\ \citenamefont
  {Kim}(2008)}]{lee_conceptual_2008}%
  \BibitemOpen
  \bibfield  {author} {\bibinfo {author} {\bibfnamefont {K.-S.}\ \bibnamefont
  {Lee}}\ and\ \bibinfo {author} {\bibfnamefont {S.-K.}\ \bibnamefont {Kim}},\
  }\href {\doibase 10.1063/1.2975235} {\bibfield  {journal} {\bibinfo
  {journal} {J. Appl. Phys.}\ }\textbf {\bibinfo {volume} {104}},\ \bibinfo
  {pages} {053909} (\bibinfo {year} {2008})}\BibitemShut {NoStop}%
\bibitem [{\citenamefont {Andreakou}\ \emph {et~al.}(2014)\citenamefont
  {Andreakou}, \citenamefont {Poltavtsev}, \citenamefont {Leonard},
  \citenamefont {Calman}, \citenamefont {Remeika}, \citenamefont {Kuznetsova},
  \citenamefont {Butov}, \citenamefont {Wilkes}, \citenamefont {Hanson},\ and\
  \citenamefont {Gossard}}]{andreakou_optically_2014}%
  \BibitemOpen
  \bibfield  {author} {\bibinfo {author} {\bibfnamefont {P.}~\bibnamefont
  {Andreakou}}, \bibinfo {author} {\bibfnamefont {S.~V.}\ \bibnamefont
  {Poltavtsev}}, \bibinfo {author} {\bibfnamefont {J.~R.}\ \bibnamefont
  {Leonard}}, \bibinfo {author} {\bibfnamefont {E.~V.}\ \bibnamefont {Calman}},
  \bibinfo {author} {\bibfnamefont {M.}~\bibnamefont {Remeika}}, \bibinfo
  {author} {\bibfnamefont {Y.~Y.}\ \bibnamefont {Kuznetsova}}, \bibinfo
  {author} {\bibfnamefont {L.~V.}\ \bibnamefont {Butov}}, \bibinfo {author}
  {\bibfnamefont {J.}~\bibnamefont {Wilkes}}, \bibinfo {author} {\bibfnamefont
  {M.}~\bibnamefont {Hanson}}, \ and\ \bibinfo {author} {\bibfnamefont {A.~C.}\
  \bibnamefont {Gossard}},\ }\href {\doibase 10.1063/1.4866855} {\bibfield
  {journal} {\bibinfo  {journal} {Appl. Phys. Lett.}\ }\textbf {\bibinfo
  {volume} {104}},\ \bibinfo {pages} {091101} (\bibinfo {year}
  {2014})}\BibitemShut {NoStop}%
\bibitem [{\citenamefont {Feng}\ \emph {et~al.}(2005)\citenamefont {Feng},
  \citenamefont {Holonyak}, \citenamefont {Walter},\ and\ \citenamefont
  {Chan}}]{feng_room_2005}%
  \BibitemOpen
  \bibfield  {author} {\bibinfo {author} {\bibfnamefont {M.}~\bibnamefont
  {Feng}}, \bibinfo {author} {\bibfnamefont {N.}~\bibnamefont {Holonyak}},
  \bibinfo {author} {\bibfnamefont {G.}~\bibnamefont {Walter}}, \ and\ \bibinfo
  {author} {\bibfnamefont {R.}~\bibnamefont {Chan}},\ }\href {\doibase
  10.1063/1.2058213} {\bibfield  {journal} {\bibinfo  {journal} {Appl. Phys.
  Lett.}\ }\textbf {\bibinfo {volume} {87}},\ \bibinfo {pages} {131103}
  (\bibinfo {year} {2005})}\BibitemShut {NoStop}%
\bibitem [{\citenamefont {{Currivan-Incorvia}}\ \emph
  {et~al.}(2016)\citenamefont {{Currivan-Incorvia}}, \citenamefont {Siddiqui},
  \citenamefont {Dutta}, \citenamefont {Evarts}, \citenamefont {Zhang},
  \citenamefont {Bono}, \citenamefont {Ross},\ and\ \citenamefont
  {Baldo}}]{currivan-incorvia_logic_2016}%
  \BibitemOpen
  \bibfield  {author} {\bibinfo {author} {\bibfnamefont {J.~A.}\ \bibnamefont
  {{Currivan-Incorvia}}}, \bibinfo {author} {\bibfnamefont {S.}~\bibnamefont
  {Siddiqui}}, \bibinfo {author} {\bibfnamefont {S.}~\bibnamefont {Dutta}},
  \bibinfo {author} {\bibfnamefont {E.~R.}\ \bibnamefont {Evarts}}, \bibinfo
  {author} {\bibfnamefont {J.}~\bibnamefont {Zhang}}, \bibinfo {author}
  {\bibfnamefont {D.}~\bibnamefont {Bono}}, \bibinfo {author} {\bibfnamefont
  {C.~A.}\ \bibnamefont {Ross}}, \ and\ \bibinfo {author} {\bibfnamefont
  {M.~A.}\ \bibnamefont {Baldo}},\ }\href {\doibase 10.1038/ncomms10275}
  {\bibfield  {journal} {\bibinfo  {journal} {Nat. Commun.}\ }\textbf {\bibinfo
  {volume} {7}},\ \bibinfo {pages} {10275} (\bibinfo {year}
  {2016})}\BibitemShut {NoStop}%
\bibitem [{\citenamefont {Pino}\ \emph {et~al.}(2021)\citenamefont {Pino},
  \citenamefont {Dreiling}, \citenamefont {Figgatt}, \citenamefont {Gaebler},
  \citenamefont {Moses}, \citenamefont {Allman}, \citenamefont {Baldwin},
  \citenamefont {{Foss-Feig}}, \citenamefont {Hayes}, \citenamefont {Mayer},
  \citenamefont {{Ryan-Anderson}},\ and\ \citenamefont
  {Neyenhuis}}]{pino_demonstration_2021}%
  \BibitemOpen
  \bibfield  {author} {\bibinfo {author} {\bibfnamefont {J.~M.}\ \bibnamefont
  {Pino}}, \bibinfo {author} {\bibfnamefont {J.~M.}\ \bibnamefont {Dreiling}},
  \bibinfo {author} {\bibfnamefont {C.}~\bibnamefont {Figgatt}}, \bibinfo
  {author} {\bibfnamefont {J.~P.}\ \bibnamefont {Gaebler}}, \bibinfo {author}
  {\bibfnamefont {S.~A.}\ \bibnamefont {Moses}}, \bibinfo {author}
  {\bibfnamefont {M.~S.}\ \bibnamefont {Allman}}, \bibinfo {author}
  {\bibfnamefont {C.~H.}\ \bibnamefont {Baldwin}}, \bibinfo {author}
  {\bibfnamefont {M.}~\bibnamefont {{Foss-Feig}}}, \bibinfo {author}
  {\bibfnamefont {D.}~\bibnamefont {Hayes}}, \bibinfo {author} {\bibfnamefont
  {K.}~\bibnamefont {Mayer}}, \bibinfo {author} {\bibfnamefont
  {C.}~\bibnamefont {{Ryan-Anderson}}}, \ and\ \bibinfo {author} {\bibfnamefont
  {B.}~\bibnamefont {Neyenhuis}},\ }\href {\doibase 10.1038/s41586-021-03318-4}
  {\bibfield  {journal} {\bibinfo  {journal} {Nature}\ }\textbf {\bibinfo
  {volume} {592}},\ \bibinfo {pages} {209} (\bibinfo {year}
  {2021})}\BibitemShut {NoStop}%
\bibitem [{\citenamefont {Kranzl}\ \emph {et~al.}(2022)\citenamefont {Kranzl},
  \citenamefont {Joshi}, \citenamefont {Maier}, \citenamefont {Brydges},
  \citenamefont {Franke}, \citenamefont {Blatt},\ and\ \citenamefont
  {Roos}}]{kranzl_controlling_2022}%
  \BibitemOpen
  \bibfield  {author} {\bibinfo {author} {\bibfnamefont {F.}~\bibnamefont
  {Kranzl}}, \bibinfo {author} {\bibfnamefont {M.~K.}\ \bibnamefont {Joshi}},
  \bibinfo {author} {\bibfnamefont {C.}~\bibnamefont {Maier}}, \bibinfo
  {author} {\bibfnamefont {T.}~\bibnamefont {Brydges}}, \bibinfo {author}
  {\bibfnamefont {J.}~\bibnamefont {Franke}}, \bibinfo {author} {\bibfnamefont
  {R.}~\bibnamefont {Blatt}}, \ and\ \bibinfo {author} {\bibfnamefont {C.~F.}\
  \bibnamefont {Roos}},\ }\href {\doibase 10.1103/PhysRevA.105.052426}
  {\bibfield  {journal} {\bibinfo  {journal} {Phys. Rev. A}\ }\textbf {\bibinfo
  {volume} {105}},\ \bibinfo {pages} {052426} (\bibinfo {year}
  {2022})}\BibitemShut {NoStop}%
\bibitem [{\citenamefont {Egan}\ \emph {et~al.}(2021)\citenamefont {Egan},
  \citenamefont {Debroy}, \citenamefont {Noel}, \citenamefont {Risinger},
  \citenamefont {Zhu}, \citenamefont {Biswas}, \citenamefont {Newman},
  \citenamefont {Li}, \citenamefont {Brown}, \citenamefont {Cetina},\ and\
  \citenamefont {Monroe}}]{egan_fault-tolerant_2021}%
  \BibitemOpen
  \bibfield  {author} {\bibinfo {author} {\bibfnamefont {L.}~\bibnamefont
  {Egan}}, \bibinfo {author} {\bibfnamefont {D.~M.}\ \bibnamefont {Debroy}},
  \bibinfo {author} {\bibfnamefont {C.}~\bibnamefont {Noel}}, \bibinfo {author}
  {\bibfnamefont {A.}~\bibnamefont {Risinger}}, \bibinfo {author}
  {\bibfnamefont {D.}~\bibnamefont {Zhu}}, \bibinfo {author} {\bibfnamefont
  {D.}~\bibnamefont {Biswas}}, \bibinfo {author} {\bibfnamefont
  {M.}~\bibnamefont {Newman}}, \bibinfo {author} {\bibfnamefont
  {M.}~\bibnamefont {Li}}, \bibinfo {author} {\bibfnamefont {K.~R.}\
  \bibnamefont {Brown}}, \bibinfo {author} {\bibfnamefont {M.}~\bibnamefont
  {Cetina}}, \ and\ \bibinfo {author} {\bibfnamefont {C.}~\bibnamefont
  {Monroe}},\ }\href {\doibase 10.1038/s41586-021-03928-y} {\bibfield
  {journal} {\bibinfo  {journal} {Nature}\ }\textbf {\bibinfo {volume} {598}},\
  \bibinfo {pages} {281} (\bibinfo {year} {2021})}\BibitemShut {NoStop}%
\bibitem [{\citenamefont {Postler}\ \emph {et~al.}(2022)\citenamefont
  {Postler}, \citenamefont {Heu{$\beta$}en}, \citenamefont {Pogorelov},
  \citenamefont {Rispler}, \citenamefont {Feldker}, \citenamefont {Meth},
  \citenamefont {Marciniak}, \citenamefont {Stricker}, \citenamefont
  {Ringbauer}, \citenamefont {Blatt}, \citenamefont {Schindler}, \citenamefont
  {M{\"u}ller},\ and\ \citenamefont {Monz}}]{postler_demonstration_2022}%
  \BibitemOpen
  \bibfield  {author} {\bibinfo {author} {\bibfnamefont {L.}~\bibnamefont
  {Postler}}, \bibinfo {author} {\bibfnamefont {S.}~\bibnamefont
  {Heu{$\beta$}en}}, \bibinfo {author} {\bibfnamefont {I.}~\bibnamefont
  {Pogorelov}}, \bibinfo {author} {\bibfnamefont {M.}~\bibnamefont {Rispler}},
  \bibinfo {author} {\bibfnamefont {T.}~\bibnamefont {Feldker}}, \bibinfo
  {author} {\bibfnamefont {M.}~\bibnamefont {Meth}}, \bibinfo {author}
  {\bibfnamefont {C.~D.}\ \bibnamefont {Marciniak}}, \bibinfo {author}
  {\bibfnamefont {R.}~\bibnamefont {Stricker}}, \bibinfo {author}
  {\bibfnamefont {M.}~\bibnamefont {Ringbauer}}, \bibinfo {author}
  {\bibfnamefont {R.}~\bibnamefont {Blatt}}, \bibinfo {author} {\bibfnamefont
  {P.}~\bibnamefont {Schindler}}, \bibinfo {author} {\bibfnamefont
  {M.}~\bibnamefont {M{\"u}ller}}, \ and\ \bibinfo {author} {\bibfnamefont
  {T.}~\bibnamefont {Monz}},\ }\href {\doibase 10.1038/s41586-022-04721-1}
  {\bibfield  {journal} {\bibinfo  {journal} {Nature}\ }\textbf {\bibinfo
  {volume} {605}},\ \bibinfo {pages} {675} (\bibinfo {year}
  {2022})}\BibitemShut {NoStop}%
\bibitem [{\citenamefont {Herrmann}\ \emph {et~al.}(2022)\citenamefont
  {Herrmann}, \citenamefont {Llima}, \citenamefont {Remm}, \citenamefont
  {Zapletal}, \citenamefont {McMahon}, \citenamefont {Scarato}, \citenamefont
  {Swiadek}, \citenamefont {Andersen}, \citenamefont {Hellings}, \citenamefont
  {Krinner} \emph {et~al.}}]{herrmann_realizing_2022}%
  \BibitemOpen
  \bibfield  {author} {\bibinfo {author} {\bibfnamefont {J.}~\bibnamefont
  {Herrmann}}, \bibinfo {author} {\bibfnamefont {S.~M.}\ \bibnamefont {Llima}},
  \bibinfo {author} {\bibfnamefont {A.}~\bibnamefont {Remm}}, \bibinfo {author}
  {\bibfnamefont {P.}~\bibnamefont {Zapletal}}, \bibinfo {author}
  {\bibfnamefont {N.~A.}\ \bibnamefont {McMahon}}, \bibinfo {author}
  {\bibfnamefont {C.}~\bibnamefont {Scarato}}, \bibinfo {author} {\bibfnamefont
  {F.}~\bibnamefont {Swiadek}}, \bibinfo {author} {\bibfnamefont {C.~K.}\
  \bibnamefont {Andersen}}, \bibinfo {author} {\bibfnamefont {C.}~\bibnamefont
  {Hellings}}, \bibinfo {author} {\bibfnamefont {S.}~\bibnamefont {Krinner}},
  \emph {et~al.},\ }\href {\doibase 10.1038/s41467-022-31679-5} {\bibfield
  {journal} {\bibinfo  {journal} {Nat. Commun.}\ }\textbf {\bibinfo {volume}
  {13}},\ \bibinfo {pages} {4144} (\bibinfo {year} {2022})}\BibitemShut
  {NoStop}%
\bibitem [{\citenamefont {Chen}\ \emph {et~al.}(2021)\citenamefont {Chen},
  \citenamefont {Satzinger}, \citenamefont {Atalaya}, \citenamefont {Korotkov},
  \citenamefont {Dunsworth}, \citenamefont {Sank}, \citenamefont {Quintana},
  \citenamefont {McEwen}, \citenamefont {Barends}, \citenamefont {Klimov} \emph
  {et~al.}}]{google_quantum_ai_exponential_2021}%
  \BibitemOpen
  \bibfield  {author} {\bibinfo {author} {\bibfnamefont {Z.}~\bibnamefont
  {Chen}}, \bibinfo {author} {\bibfnamefont {K.~J.}\ \bibnamefont {Satzinger}},
  \bibinfo {author} {\bibfnamefont {J.}~\bibnamefont {Atalaya}}, \bibinfo
  {author} {\bibfnamefont {A.~N.}\ \bibnamefont {Korotkov}}, \bibinfo {author}
  {\bibfnamefont {A.}~\bibnamefont {Dunsworth}}, \bibinfo {author}
  {\bibfnamefont {D.}~\bibnamefont {Sank}}, \bibinfo {author} {\bibfnamefont
  {C.}~\bibnamefont {Quintana}}, \bibinfo {author} {\bibfnamefont
  {M.}~\bibnamefont {McEwen}}, \bibinfo {author} {\bibfnamefont
  {R.}~\bibnamefont {Barends}}, \bibinfo {author} {\bibfnamefont {P.~V.}\
  \bibnamefont {Klimov}},  \emph {et~al.},\ }\href {\doibase
  10.1038/s41586-021-03588-y} {\bibfield  {journal} {\bibinfo  {journal}
  {Nature}\ }\textbf {\bibinfo {volume} {595}},\ \bibinfo {pages} {383}
  (\bibinfo {year} {2021})}\BibitemShut {NoStop}%
\bibitem [{\citenamefont {Olsacher}\ \emph {et~al.}(2020)\citenamefont
  {Olsacher}, \citenamefont {Postler}, \citenamefont {Schindler}, \citenamefont
  {Monz}, \citenamefont {Zoller},\ and\ \citenamefont
  {Sieberer}}]{olsacher_scalable_2020}%
  \BibitemOpen
  \bibfield  {author} {\bibinfo {author} {\bibfnamefont {T.}~\bibnamefont
  {Olsacher}}, \bibinfo {author} {\bibfnamefont {L.}~\bibnamefont {Postler}},
  \bibinfo {author} {\bibfnamefont {P.}~\bibnamefont {Schindler}}, \bibinfo
  {author} {\bibfnamefont {T.}~\bibnamefont {Monz}}, \bibinfo {author}
  {\bibfnamefont {P.}~\bibnamefont {Zoller}}, \ and\ \bibinfo {author}
  {\bibfnamefont {L.~M.}\ \bibnamefont {Sieberer}},\ }\href {\doibase
  10.1103/PRXQuantum.1.020316} {\bibfield  {journal} {\bibinfo  {journal} {PRX
  Quantum}\ }\textbf {\bibinfo {volume} {1}},\ \bibinfo {pages} {020316}
  (\bibinfo {year} {2020})}\BibitemShut {NoStop}%
\bibitem [{\citenamefont {Zhang}\ \emph {et~al.}(2020)\citenamefont {Zhang},
  \citenamefont {Pokorny}, \citenamefont {Li}, \citenamefont {Higgins},
  \citenamefont {P{\"o}schl}, \citenamefont {Lesanovsky},\ and\ \citenamefont
  {Hennrich}}]{zhang_submicrosecond_2020}%
  \BibitemOpen
  \bibfield  {author} {\bibinfo {author} {\bibfnamefont {C.}~\bibnamefont
  {Zhang}}, \bibinfo {author} {\bibfnamefont {F.}~\bibnamefont {Pokorny}},
  \bibinfo {author} {\bibfnamefont {W.}~\bibnamefont {Li}}, \bibinfo {author}
  {\bibfnamefont {G.}~\bibnamefont {Higgins}}, \bibinfo {author} {\bibfnamefont
  {A.}~\bibnamefont {P{\"o}schl}}, \bibinfo {author} {\bibfnamefont
  {I.}~\bibnamefont {Lesanovsky}}, \ and\ \bibinfo {author} {\bibfnamefont
  {M.}~\bibnamefont {Hennrich}},\ }\href {\doibase 10.1038/s41586-020-2152-9}
  {\bibfield  {journal} {\bibinfo  {journal} {Nature}\ }\textbf {\bibinfo
  {volume} {580}},\ \bibinfo {pages} {345} (\bibinfo {year}
  {2020})}\BibitemShut {NoStop}%
\bibitem [{\citenamefont {Krinner}\ \emph {et~al.}(2022)\citenamefont
  {Krinner}, \citenamefont {Lacroix}, \citenamefont {Remm}, \citenamefont
  {Di~Paolo}, \citenamefont {Genois}, \citenamefont {Leroux}, \citenamefont
  {Hellings}, \citenamefont {Lazar}, \citenamefont {Swiadek}, \citenamefont
  {Herrmann} \emph {et~al.}}]{krinner_realizing_2022}%
  \BibitemOpen
  \bibfield  {author} {\bibinfo {author} {\bibfnamefont {S.}~\bibnamefont
  {Krinner}}, \bibinfo {author} {\bibfnamefont {N.}~\bibnamefont {Lacroix}},
  \bibinfo {author} {\bibfnamefont {A.}~\bibnamefont {Remm}}, \bibinfo {author}
  {\bibfnamefont {A.}~\bibnamefont {Di~Paolo}}, \bibinfo {author}
  {\bibfnamefont {E.}~\bibnamefont {Genois}}, \bibinfo {author} {\bibfnamefont
  {C.}~\bibnamefont {Leroux}}, \bibinfo {author} {\bibfnamefont
  {C.}~\bibnamefont {Hellings}}, \bibinfo {author} {\bibfnamefont
  {S.}~\bibnamefont {Lazar}}, \bibinfo {author} {\bibfnamefont
  {F.}~\bibnamefont {Swiadek}}, \bibinfo {author} {\bibfnamefont
  {J.}~\bibnamefont {Herrmann}},  \emph {et~al.},\ }\href {\doibase
  10.1038/s41586-022-04566-8} {\bibfield  {journal} {\bibinfo  {journal}
  {Nature}\ }\textbf {\bibinfo {volume} {605}},\ \bibinfo {pages} {669}
  (\bibinfo {year} {2022})}\BibitemShut {NoStop}%
\bibitem [{\citenamefont {Hrmo}\ \emph {et~al.}(2022)\citenamefont {Hrmo},
  \citenamefont {Wilhelm}, \citenamefont {Gerster}, \citenamefont {{van
  Mourik}}, \citenamefont {Huber}, \citenamefont {Blatt}, \citenamefont
  {Schindler}, \citenamefont {Monz},\ and\ \citenamefont
  {Ringbauer}}]{hrmo_native_2022}%
  \BibitemOpen
  \bibfield  {author} {\bibinfo {author} {\bibfnamefont {P.}~\bibnamefont
  {Hrmo}}, \bibinfo {author} {\bibfnamefont {B.}~\bibnamefont {Wilhelm}},
  \bibinfo {author} {\bibfnamefont {L.}~\bibnamefont {Gerster}}, \bibinfo
  {author} {\bibfnamefont {M.~W.}\ \bibnamefont {{van Mourik}}}, \bibinfo
  {author} {\bibfnamefont {M.}~\bibnamefont {Huber}}, \bibinfo {author}
  {\bibfnamefont {R.}~\bibnamefont {Blatt}}, \bibinfo {author} {\bibfnamefont
  {P.}~\bibnamefont {Schindler}}, \bibinfo {author} {\bibfnamefont
  {T.}~\bibnamefont {Monz}}, \ and\ \bibinfo {author} {\bibfnamefont
  {M.}~\bibnamefont {Ringbauer}},\ }\href {https://arxiv.org/abs/2206.04104}
  {\bibfield  {journal} {\bibinfo  {journal} {arXiv:2206.04104}\ } (\bibinfo
  {year} {2022})}\BibitemShut {NoStop}%
\bibitem [{\citenamefont {Chu}\ \emph {et~al.}(2023)\citenamefont {Chu},
  \citenamefont {He}, \citenamefont {Zhou}, \citenamefont {Yuan}, \citenamefont
  {Zhang}, \citenamefont {Guo}, \citenamefont {Hai}, \citenamefont {Han},
  \citenamefont {Hu}, \citenamefont {Huang} \emph
  {et~al.}}]{chu_scalable_2023}%
  \BibitemOpen
  \bibfield  {author} {\bibinfo {author} {\bibfnamefont {J.}~\bibnamefont
  {Chu}}, \bibinfo {author} {\bibfnamefont {X.}~\bibnamefont {He}}, \bibinfo
  {author} {\bibfnamefont {Y.}~\bibnamefont {Zhou}}, \bibinfo {author}
  {\bibfnamefont {J.}~\bibnamefont {Yuan}}, \bibinfo {author} {\bibfnamefont
  {L.}~\bibnamefont {Zhang}}, \bibinfo {author} {\bibfnamefont
  {Q.}~\bibnamefont {Guo}}, \bibinfo {author} {\bibfnamefont {Y.}~\bibnamefont
  {Hai}}, \bibinfo {author} {\bibfnamefont {Z.}~\bibnamefont {Han}}, \bibinfo
  {author} {\bibfnamefont {C.-K.}\ \bibnamefont {Hu}}, \bibinfo {author}
  {\bibfnamefont {W.}~\bibnamefont {Huang}},  \emph {et~al.},\ }\href {\doibase
  10.1038/s41567-022-01813-7} {\bibfield  {journal} {\bibinfo  {journal} {Nat.
  Phys.}\ }\textbf {\bibinfo {volume} {19}},\ \bibinfo {pages} {126} (\bibinfo
  {year} {2023})}\BibitemShut {NoStop}%
\bibitem [{\citenamefont {Preskill}(2018)}]{preskill_quantum_2018}%
  \BibitemOpen
  \bibfield  {author} {\bibinfo {author} {\bibfnamefont {J.}~\bibnamefont
  {Preskill}},\ }\href {\doibase 10.22331/q-2018-08-06-79} {\bibfield
  {journal} {\bibinfo  {journal} {Quantum}\ }\textbf {\bibinfo {volume} {2}},\
  \bibinfo {pages} {79} (\bibinfo {year} {2018})}\BibitemShut {NoStop}%
\bibitem [{\citenamefont {Bharti}\ \emph {et~al.}(2022)\citenamefont {Bharti},
  \citenamefont {{Cervera-Lierta}}, \citenamefont {Kyaw}, \citenamefont {Haug},
  \citenamefont {{Alperin-Lea}}, \citenamefont {Anand}, \citenamefont
  {Degroote}, \citenamefont {Heimonen}, \citenamefont {Kottmann}, \citenamefont
  {Menke} \emph {et~al.}}]{bharti_noisy_2022}%
  \BibitemOpen
  \bibfield  {author} {\bibinfo {author} {\bibfnamefont {K.}~\bibnamefont
  {Bharti}}, \bibinfo {author} {\bibfnamefont {A.}~\bibnamefont
  {{Cervera-Lierta}}}, \bibinfo {author} {\bibfnamefont {T.~H.}\ \bibnamefont
  {Kyaw}}, \bibinfo {author} {\bibfnamefont {T.}~\bibnamefont {Haug}}, \bibinfo
  {author} {\bibfnamefont {S.}~\bibnamefont {{Alperin-Lea}}}, \bibinfo {author}
  {\bibfnamefont {A.}~\bibnamefont {Anand}}, \bibinfo {author} {\bibfnamefont
  {M.}~\bibnamefont {Degroote}}, \bibinfo {author} {\bibfnamefont
  {H.}~\bibnamefont {Heimonen}}, \bibinfo {author} {\bibfnamefont {J.~S.}\
  \bibnamefont {Kottmann}}, \bibinfo {author} {\bibfnamefont {T.}~\bibnamefont
  {Menke}},  \emph {et~al.},\ }\href {\doibase 10.1103/RevModPhys.94.015004}
  {\bibfield  {journal} {\bibinfo  {journal} {Rev. Mod. Phys.}\ }\textbf
  {\bibinfo {volume} {94}},\ \bibinfo {pages} {015004} (\bibinfo {year}
  {2022})}\BibitemShut {NoStop}%
\bibitem [{\citenamefont {Foldager}\ \emph {et~al.}(2022)\citenamefont
  {Foldager}, \citenamefont {Pesah},\ and\ \citenamefont
  {Hansen}}]{foldager_noise-assisted_2022}%
  \BibitemOpen
  \bibfield  {author} {\bibinfo {author} {\bibfnamefont {J.}~\bibnamefont
  {Foldager}}, \bibinfo {author} {\bibfnamefont {A.}~\bibnamefont {Pesah}}, \
  and\ \bibinfo {author} {\bibfnamefont {L.~K.}\ \bibnamefont {Hansen}},\
  }\href {\doibase 10.1038/s41598-022-07296-z} {\bibfield  {journal} {\bibinfo
  {journal} {Sci. Rep.}\ }\textbf {\bibinfo {volume} {12}},\ \bibinfo {pages}
  {3862} (\bibinfo {year} {2022})}\BibitemShut {NoStop}%
\bibitem [{\citenamefont {Wu}\ and\ \citenamefont
  {Hsieh}(2019)}]{wu_variational_2019}%
  \BibitemOpen
  \bibfield  {author} {\bibinfo {author} {\bibfnamefont {J.}~\bibnamefont
  {Wu}}\ and\ \bibinfo {author} {\bibfnamefont {T.~H.}\ \bibnamefont {Hsieh}},\
  }\href {\doibase 10.1103/PhysRevLett.123.220502} {\bibfield  {journal}
  {\bibinfo  {journal} {Phys. Rev. Lett.}\ }\textbf {\bibinfo {volume} {123}},\
  \bibinfo {pages} {220502} (\bibinfo {year} {2019})}\BibitemShut {NoStop}%
\bibitem [{\citenamefont {Zhu}\ \emph {et~al.}(2020)\citenamefont {Zhu},
  \citenamefont {Johri}, \citenamefont {Linke}, \citenamefont {Landsman},
  \citenamefont {Huerta~Alderete}, \citenamefont {Nguyen}, \citenamefont
  {Matsuura}, \citenamefont {Hsieh},\ and\ \citenamefont
  {Monroe}}]{zhu_generation_2020}%
  \BibitemOpen
  \bibfield  {author} {\bibinfo {author} {\bibfnamefont {D.}~\bibnamefont
  {Zhu}}, \bibinfo {author} {\bibfnamefont {S.}~\bibnamefont {Johri}}, \bibinfo
  {author} {\bibfnamefont {N.~M.}\ \bibnamefont {Linke}}, \bibinfo {author}
  {\bibfnamefont {K.~A.}\ \bibnamefont {Landsman}}, \bibinfo {author}
  {\bibfnamefont {C.}~\bibnamefont {Huerta~Alderete}}, \bibinfo {author}
  {\bibfnamefont {N.~H.}\ \bibnamefont {Nguyen}}, \bibinfo {author}
  {\bibfnamefont {A.~Y.}\ \bibnamefont {Matsuura}}, \bibinfo {author}
  {\bibfnamefont {T.~H.}\ \bibnamefont {Hsieh}}, \ and\ \bibinfo {author}
  {\bibfnamefont {C.}~\bibnamefont {Monroe}},\ }\href {\doibase
  10.1073/pnas.2006337117} {\bibfield  {journal} {\bibinfo  {journal}
  {Proceedings of the National Academy of Sciences}\ }\textbf {\bibinfo
  {volume} {117}},\ \bibinfo {pages} {25402} (\bibinfo {year}
  {2020})}\BibitemShut {NoStop}%
\bibitem [{\citenamefont {Wang}\ \emph {et~al.}(2021)\citenamefont {Wang},
  \citenamefont {Li},\ and\ \citenamefont {Wang}}]{wang_variational_2021}%
  \BibitemOpen
  \bibfield  {author} {\bibinfo {author} {\bibfnamefont {Y.}~\bibnamefont
  {Wang}}, \bibinfo {author} {\bibfnamefont {G.}~\bibnamefont {Li}}, \ and\
  \bibinfo {author} {\bibfnamefont {X.}~\bibnamefont {Wang}},\ }\href {\doibase
  10.1103/PhysRevApplied.16.054035} {\bibfield  {journal} {\bibinfo  {journal}
  {Phys. Rev. Appl}\ }\textbf {\bibinfo {volume} {16}},\ \bibinfo {pages}
  {054035} (\bibinfo {year} {2021})}\BibitemShut {NoStop}%
\bibitem [{\citenamefont {Mazzola}\ \emph {et~al.}(2019)\citenamefont
  {Mazzola}, \citenamefont {Ollitrault}, \citenamefont {Barkoutsos},\ and\
  \citenamefont {Tavernelli}}]{mazzola_nonunitary_2019}%
  \BibitemOpen
  \bibfield  {author} {\bibinfo {author} {\bibfnamefont {G.}~\bibnamefont
  {Mazzola}}, \bibinfo {author} {\bibfnamefont {P.~J.}\ \bibnamefont
  {Ollitrault}}, \bibinfo {author} {\bibfnamefont {P.~K.}\ \bibnamefont
  {Barkoutsos}}, \ and\ \bibinfo {author} {\bibfnamefont {I.}~\bibnamefont
  {Tavernelli}},\ }\href {\doibase 10.1103/PhysRevLett.123.130501} {\bibfield
  {journal} {\bibinfo  {journal} {Phys. Rev. Lett.}\ }\textbf {\bibinfo
  {volume} {123}},\ \bibinfo {pages} {130501} (\bibinfo {year}
  {2019})}\BibitemShut {NoStop}%
\bibitem [{\citenamefont {Verdon}\ \emph {et~al.}(2019)\citenamefont {Verdon},
  \citenamefont {Marks}, \citenamefont {Nanda}, \citenamefont {Leichenauer},\
  and\ \citenamefont {Hidary}}]{verdon_quantum_2019}%
  \BibitemOpen
  \bibfield  {author} {\bibinfo {author} {\bibfnamefont {G.}~\bibnamefont
  {Verdon}}, \bibinfo {author} {\bibfnamefont {J.}~\bibnamefont {Marks}},
  \bibinfo {author} {\bibfnamefont {S.}~\bibnamefont {Nanda}}, \bibinfo
  {author} {\bibfnamefont {S.}~\bibnamefont {Leichenauer}}, \ and\ \bibinfo
  {author} {\bibfnamefont {J.}~\bibnamefont {Hidary}},\ }\href
  {http://arxiv.org/abs/1910.02071} {\bibfield  {journal} {\bibinfo  {journal}
  {arXiv:1910.02071}\ } (\bibinfo {year} {2019})}\BibitemShut {NoStop}%
\bibitem [{\citenamefont {Wang}\ \emph {et~al.}(2011)\citenamefont {Wang},
  \citenamefont {Ashhab},\ and\ \citenamefont {Nori}}]{wang_quantum_2011}%
  \BibitemOpen
  \bibfield  {author} {\bibinfo {author} {\bibfnamefont {H.}~\bibnamefont
  {Wang}}, \bibinfo {author} {\bibfnamefont {S.}~\bibnamefont {Ashhab}}, \ and\
  \bibinfo {author} {\bibfnamefont {F.}~\bibnamefont {Nori}},\ }\href {\doibase
  10.1103/PhysRevA.83.062317} {\bibfield  {journal} {\bibinfo  {journal} {Phys.
  Rev. A}\ }\textbf {\bibinfo {volume} {83}},\ \bibinfo {pages} {062317}
  (\bibinfo {year} {2011})}\BibitemShut {NoStop}%
\bibitem [{\citenamefont {Del~Re}\ \emph {et~al.}(2020)\citenamefont {Del~Re},
  \citenamefont {Rost}, \citenamefont {Kemper},\ and\ \citenamefont
  {Freericks}}]{del_re_driven-dissipative_2020}%
  \BibitemOpen
  \bibfield  {author} {\bibinfo {author} {\bibfnamefont {L.}~\bibnamefont
  {Del~Re}}, \bibinfo {author} {\bibfnamefont {B.}~\bibnamefont {Rost}},
  \bibinfo {author} {\bibfnamefont {A.~F.}\ \bibnamefont {Kemper}}, \ and\
  \bibinfo {author} {\bibfnamefont {J.~K.}\ \bibnamefont {Freericks}},\ }\href
  {\doibase 10.1103/PhysRevB.102.125112} {\bibfield  {journal} {\bibinfo
  {journal} {Phys. Rev. B}\ }\textbf {\bibinfo {volume} {102}},\ \bibinfo
  {pages} {125112} (\bibinfo {year} {2020})}\BibitemShut {NoStop}%
\bibitem [{\citenamefont {Hu}\ \emph {et~al.}(2020)\citenamefont {Hu},
  \citenamefont {Xia},\ and\ \citenamefont {Kais}}]{hu_quantum_2020}%
  \BibitemOpen
  \bibfield  {author} {\bibinfo {author} {\bibfnamefont {Z.}~\bibnamefont
  {Hu}}, \bibinfo {author} {\bibfnamefont {R.}~\bibnamefont {Xia}}, \ and\
  \bibinfo {author} {\bibfnamefont {S.}~\bibnamefont {Kais}},\ }\href {\doibase
  10.1038/s41598-020-60321-x} {\bibfield  {journal} {\bibinfo  {journal} {Sci.
  Rep.}\ }\textbf {\bibinfo {volume} {10}},\ \bibinfo {pages} {3301} (\bibinfo
  {year} {2020})}\BibitemShut {NoStop}%
\bibitem [{\citenamefont {Hu}\ \emph {et~al.}(2022)\citenamefont {Hu},
  \citenamefont {{Head-Marsden}}, \citenamefont {Mazziotti}, \citenamefont
  {Narang},\ and\ \citenamefont {Kais}}]{hu_general_2022}%
  \BibitemOpen
  \bibfield  {author} {\bibinfo {author} {\bibfnamefont {Z.}~\bibnamefont
  {Hu}}, \bibinfo {author} {\bibfnamefont {K.}~\bibnamefont {{Head-Marsden}}},
  \bibinfo {author} {\bibfnamefont {D.~A.}\ \bibnamefont {Mazziotti}}, \bibinfo
  {author} {\bibfnamefont {P.}~\bibnamefont {Narang}}, \ and\ \bibinfo {author}
  {\bibfnamefont {S.}~\bibnamefont {Kais}},\ }\href {\doibase
  10.22331/q-2022-05-30-726} {\bibfield  {journal} {\bibinfo  {journal}
  {Quantum}\ }\textbf {\bibinfo {volume} {6}},\ \bibinfo {pages} {726}
  (\bibinfo {year} {2022})}\BibitemShut {NoStop}%
\bibitem [{\citenamefont {Lepp{\"a}kangas}\ \emph {et~al.}(2022)\citenamefont
  {Lepp{\"a}kangas}, \citenamefont {Vogt}, \citenamefont {Fratus},
  \citenamefont {Bark}, \citenamefont {Vaitkus}, \citenamefont {Stadler},
  \citenamefont {Reiner}, \citenamefont {Zanker},\ and\ \citenamefont
  {Marthaler}}]{leppakangas_quantum_2022}%
  \BibitemOpen
  \bibfield  {author} {\bibinfo {author} {\bibfnamefont {J.}~\bibnamefont
  {Lepp{\"a}kangas}}, \bibinfo {author} {\bibfnamefont {N.}~\bibnamefont
  {Vogt}}, \bibinfo {author} {\bibfnamefont {K.~R.}\ \bibnamefont {Fratus}},
  \bibinfo {author} {\bibfnamefont {K.}~\bibnamefont {Bark}}, \bibinfo {author}
  {\bibfnamefont {J.~A.}\ \bibnamefont {Vaitkus}}, \bibinfo {author}
  {\bibfnamefont {P.}~\bibnamefont {Stadler}}, \bibinfo {author} {\bibfnamefont
  {J.-M.}\ \bibnamefont {Reiner}}, \bibinfo {author} {\bibfnamefont
  {S.}~\bibnamefont {Zanker}}, \ and\ \bibinfo {author} {\bibfnamefont
  {M.}~\bibnamefont {Marthaler}},\ }\href {https://arxiv.org/abs/2210.12138}
  {\bibfield  {journal} {\bibinfo  {journal} {arXiv:2210.12138}\ } (\bibinfo
  {year} {2022})}\BibitemShut {NoStop}%
\bibitem [{\citenamefont {Cong}\ \emph {et~al.}(2019)\citenamefont {Cong},
  \citenamefont {Choi},\ and\ \citenamefont {Lukin}}]{cong_quantum_2019}%
  \BibitemOpen
  \bibfield  {author} {\bibinfo {author} {\bibfnamefont {I.}~\bibnamefont
  {Cong}}, \bibinfo {author} {\bibfnamefont {S.}~\bibnamefont {Choi}}, \ and\
  \bibinfo {author} {\bibfnamefont {M.~D.}\ \bibnamefont {Lukin}},\ }\href
  {\doibase 10.1038/s41567-019-0648-8} {\bibfield  {journal} {\bibinfo
  {journal} {Nat. Phys.}\ }\textbf {\bibinfo {volume} {15}},\ \bibinfo {pages}
  {1273} (\bibinfo {year} {2019})}\BibitemShut {NoStop}%
\bibitem [{\citenamefont {Harrington}\ \emph {et~al.}(2022)\citenamefont
  {Harrington}, \citenamefont {Mueller},\ and\ \citenamefont
  {Murch}}]{harrington_engineered_2022}%
  \BibitemOpen
  \bibfield  {author} {\bibinfo {author} {\bibfnamefont {P.~M.}\ \bibnamefont
  {Harrington}}, \bibinfo {author} {\bibfnamefont {E.~J.}\ \bibnamefont
  {Mueller}}, \ and\ \bibinfo {author} {\bibfnamefont {K.~W.}\ \bibnamefont
  {Murch}},\ }\href {\doibase 10.1038/s42254-022-00494-8} {\bibfield  {journal}
  {\bibinfo  {journal} {Nature Reviews Physics}\ }\textbf {\bibinfo {volume}
  {4}},\ \bibinfo {pages} {660} (\bibinfo {year} {2022})}\BibitemShut {NoStop}%
\bibitem [{\citenamefont {Kronwald}\ \emph {et~al.}(2014)\citenamefont
  {Kronwald}, \citenamefont {Marquardt},\ and\ \citenamefont
  {Clerk}}]{kronwald_dissipative_2014}%
  \BibitemOpen
  \bibfield  {author} {\bibinfo {author} {\bibfnamefont {A.}~\bibnamefont
  {Kronwald}}, \bibinfo {author} {\bibfnamefont {F.}~\bibnamefont {Marquardt}},
  \ and\ \bibinfo {author} {\bibfnamefont {A.~A.}\ \bibnamefont {Clerk}},\
  }\href {\doibase 10.1088/1367-2630/16/6/063058} {\bibfield  {journal}
  {\bibinfo  {journal} {New J. Phys.}\ }\textbf {\bibinfo {volume} {16}},\
  \bibinfo {pages} {063058} (\bibinfo {year} {2014})}\BibitemShut {NoStop}%
\bibitem [{\citenamefont {Agarwal}\ and\ \citenamefont
  {Huang}(2016)}]{agarwal_strong_2016}%
  \BibitemOpen
  \bibfield  {author} {\bibinfo {author} {\bibfnamefont {G.~S.}\ \bibnamefont
  {Agarwal}}\ and\ \bibinfo {author} {\bibfnamefont {S.}~\bibnamefont
  {Huang}},\ }\href {\doibase 10.1103/PhysRevA.93.043844} {\bibfield  {journal}
  {\bibinfo  {journal} {Phys. Rev. A}\ }\textbf {\bibinfo {volume} {93}},\
  \bibinfo {pages} {043844} (\bibinfo {year} {2016})}\BibitemShut {NoStop}%
\bibitem [{\citenamefont {Reiter}\ \emph {et~al.}(2016)\citenamefont {Reiter},
  \citenamefont {Reeb},\ and\ \citenamefont
  {S{\o}rensen}}]{reiter_scalable_2016}%
  \BibitemOpen
  \bibfield  {author} {\bibinfo {author} {\bibfnamefont {F.}~\bibnamefont
  {Reiter}}, \bibinfo {author} {\bibfnamefont {D.}~\bibnamefont {Reeb}}, \ and\
  \bibinfo {author} {\bibfnamefont {A.~S.}\ \bibnamefont {S{\o}rensen}},\
  }\href {\doibase 10.1103/PhysRevLett.117.040501} {\bibfield  {journal}
  {\bibinfo  {journal} {Phys. Rev. Lett.}\ }\textbf {\bibinfo {volume} {117}},\
  \bibinfo {pages} {040501} (\bibinfo {year} {2016})}\BibitemShut {NoStop}%
\bibitem [{\citenamefont {Kastoryano}\ \emph {et~al.}(2011)\citenamefont
  {Kastoryano}, \citenamefont {Reiter},\ and\ \citenamefont
  {S{\o}rensen}}]{kastoryano_dissipative_2011}%
  \BibitemOpen
  \bibfield  {author} {\bibinfo {author} {\bibfnamefont {M.~J.}\ \bibnamefont
  {Kastoryano}}, \bibinfo {author} {\bibfnamefont {F.}~\bibnamefont {Reiter}},
  \ and\ \bibinfo {author} {\bibfnamefont {A.~S.}\ \bibnamefont
  {S{\o}rensen}},\ }\href {\doibase 10.1103/PhysRevLett.106.090502} {\bibfield
  {journal} {\bibinfo  {journal} {Phys. Rev. Lett.}\ }\textbf {\bibinfo
  {volume} {106}},\ \bibinfo {pages} {090502} (\bibinfo {year}
  {2011})}\BibitemShut {NoStop}%
\bibitem [{\citenamefont {Lin}\ \emph {et~al.}(2013)\citenamefont {Lin},
  \citenamefont {Gaebler}, \citenamefont {Reiter}, \citenamefont {Tan},
  \citenamefont {Bowler}, \citenamefont {S{\o}rensen}, \citenamefont
  {Leibfried},\ and\ \citenamefont {Wineland}}]{lin_dissipative_2013}%
  \BibitemOpen
  \bibfield  {author} {\bibinfo {author} {\bibfnamefont {Y.}~\bibnamefont
  {Lin}}, \bibinfo {author} {\bibfnamefont {J.~P.}\ \bibnamefont {Gaebler}},
  \bibinfo {author} {\bibfnamefont {F.}~\bibnamefont {Reiter}}, \bibinfo
  {author} {\bibfnamefont {T.~R.}\ \bibnamefont {Tan}}, \bibinfo {author}
  {\bibfnamefont {R.}~\bibnamefont {Bowler}}, \bibinfo {author} {\bibfnamefont
  {A.~S.}\ \bibnamefont {S{\o}rensen}}, \bibinfo {author} {\bibfnamefont
  {D.}~\bibnamefont {Leibfried}}, \ and\ \bibinfo {author} {\bibfnamefont
  {D.~J.}\ \bibnamefont {Wineland}},\ }\href {\doibase 10.1038/nature12801}
  {\bibfield  {journal} {\bibinfo  {journal} {Nature}\ }\textbf {\bibinfo
  {volume} {504}},\ \bibinfo {pages} {415} (\bibinfo {year}
  {2013})}\BibitemShut {NoStop}%
\bibitem [{\citenamefont {Morigi}\ \emph {et~al.}(2015)\citenamefont {Morigi},
  \citenamefont {Eschner}, \citenamefont {Cormick}, \citenamefont {Lin},
  \citenamefont {Leibfried},\ and\ \citenamefont
  {Wineland}}]{morigi_dissipative_2015}%
  \BibitemOpen
  \bibfield  {author} {\bibinfo {author} {\bibfnamefont {G.}~\bibnamefont
  {Morigi}}, \bibinfo {author} {\bibfnamefont {J.}~\bibnamefont {Eschner}},
  \bibinfo {author} {\bibfnamefont {C.}~\bibnamefont {Cormick}}, \bibinfo
  {author} {\bibfnamefont {Y.}~\bibnamefont {Lin}}, \bibinfo {author}
  {\bibfnamefont {D.}~\bibnamefont {Leibfried}}, \ and\ \bibinfo {author}
  {\bibfnamefont {D.~J.}\ \bibnamefont {Wineland}},\ }\href {\doibase
  10.1103/PhysRevLett.115.200502} {\bibfield  {journal} {\bibinfo  {journal}
  {Phys. Rev. Lett.}\ }\textbf {\bibinfo {volume} {115}},\ \bibinfo {pages}
  {200502} (\bibinfo {year} {2015})}\BibitemShut {NoStop}%
\bibitem [{\citenamefont {Cole}\ \emph {et~al.}(2021)\citenamefont {Cole},
  \citenamefont {Wu}, \citenamefont {Erickson}, \citenamefont {Hou},
  \citenamefont {Wilson}, \citenamefont {Leibfried},\ and\ \citenamefont
  {Reiter}}]{cole_dissipative_2021}%
  \BibitemOpen
  \bibfield  {author} {\bibinfo {author} {\bibfnamefont {D.~C.}\ \bibnamefont
  {Cole}}, \bibinfo {author} {\bibfnamefont {J.~J.}\ \bibnamefont {Wu}},
  \bibinfo {author} {\bibfnamefont {S.~D.}\ \bibnamefont {Erickson}}, \bibinfo
  {author} {\bibfnamefont {P.-Y.}\ \bibnamefont {Hou}}, \bibinfo {author}
  {\bibfnamefont {A.~C.}\ \bibnamefont {Wilson}}, \bibinfo {author}
  {\bibfnamefont {D.}~\bibnamefont {Leibfried}}, \ and\ \bibinfo {author}
  {\bibfnamefont {F.}~\bibnamefont {Reiter}},\ }\href {\doibase
  10.1088/1367-2630/ac09c8} {\bibfield  {journal} {\bibinfo  {journal} {New J.
  Phys.}\ }\textbf {\bibinfo {volume} {23}},\ \bibinfo {pages} {073001}
  (\bibinfo {year} {2021})}\BibitemShut {NoStop}%
\bibitem [{\citenamefont {Cole}\ \emph {et~al.}(2022)\citenamefont {Cole},
  \citenamefont {Erickson}, \citenamefont {Zarantonello}, \citenamefont {Horn},
  \citenamefont {Hou}, \citenamefont {Wu}, \citenamefont {Slichter},
  \citenamefont {Reiter}, \citenamefont {Koch},\ and\ \citenamefont
  {Leibfried}}]{cole_resource-efficient_2022}%
  \BibitemOpen
  \bibfield  {author} {\bibinfo {author} {\bibfnamefont {D.~C.}\ \bibnamefont
  {Cole}}, \bibinfo {author} {\bibfnamefont {S.~D.}\ \bibnamefont {Erickson}},
  \bibinfo {author} {\bibfnamefont {G.}~\bibnamefont {Zarantonello}}, \bibinfo
  {author} {\bibfnamefont {K.~P.}\ \bibnamefont {Horn}}, \bibinfo {author}
  {\bibfnamefont {P.-Y.}\ \bibnamefont {Hou}}, \bibinfo {author} {\bibfnamefont
  {J.~J.}\ \bibnamefont {Wu}}, \bibinfo {author} {\bibfnamefont {D.~H.}\
  \bibnamefont {Slichter}}, \bibinfo {author} {\bibfnamefont {F.}~\bibnamefont
  {Reiter}}, \bibinfo {author} {\bibfnamefont {C.~P.}\ \bibnamefont {Koch}}, \
  and\ \bibinfo {author} {\bibfnamefont {D.}~\bibnamefont {Leibfried}},\ }\href
  {\doibase 10.1103/PhysRevLett.128.080502} {\bibfield  {journal} {\bibinfo
  {journal} {Phys. Rev. Lett.}\ }\textbf {\bibinfo {volume} {128}},\ \bibinfo
  {pages} {080502} (\bibinfo {year} {2022})}\BibitemShut {NoStop}%
\bibitem [{\citenamefont {Malinowski}\ \emph {et~al.}(2021)\citenamefont
  {Malinowski}, \citenamefont {Zhang}, \citenamefont {Negnevitsky},
  \citenamefont {Rojkov}, \citenamefont {Reiter}, \citenamefont {Nguyen},
  \citenamefont {Stadler}, \citenamefont {Kienzler}, \citenamefont {Mehta},\
  and\ \citenamefont {Home}}]{malinowski_generation_2021}%
  \BibitemOpen
  \bibfield  {author} {\bibinfo {author} {\bibfnamefont {M.}~\bibnamefont
  {Malinowski}}, \bibinfo {author} {\bibfnamefont {C.}~\bibnamefont {Zhang}},
  \bibinfo {author} {\bibfnamefont {V.}~\bibnamefont {Negnevitsky}}, \bibinfo
  {author} {\bibfnamefont {I.}~\bibnamefont {Rojkov}}, \bibinfo {author}
  {\bibfnamefont {F.}~\bibnamefont {Reiter}}, \bibinfo {author} {\bibfnamefont
  {T.-L.}\ \bibnamefont {Nguyen}}, \bibinfo {author} {\bibfnamefont
  {M.}~\bibnamefont {Stadler}}, \bibinfo {author} {\bibfnamefont
  {D.}~\bibnamefont {Kienzler}}, \bibinfo {author} {\bibfnamefont {K.~K.}\
  \bibnamefont {Mehta}}, \ and\ \bibinfo {author} {\bibfnamefont {J.~P.}\
  \bibnamefont {Home}},\ }\href {\doibase 10.1103/PhysRevLett.128.080503}
  {\bibfield  {journal} {\bibinfo  {journal} {Phys. Rev. Lett.}\ }\textbf
  {\bibinfo {volume} {128}},\ \bibinfo {pages} {080503} (\bibinfo {year}
  {2021})}\BibitemShut {NoStop}%
\bibitem [{\citenamefont {Doucet}\ \emph {et~al.}(2020)\citenamefont {Doucet},
  \citenamefont {Reiter}, \citenamefont {Ranzani},\ and\ \citenamefont
  {Kamal}}]{doucet_high_2020}%
  \BibitemOpen
  \bibfield  {author} {\bibinfo {author} {\bibfnamefont {E.}~\bibnamefont
  {Doucet}}, \bibinfo {author} {\bibfnamefont {F.}~\bibnamefont {Reiter}},
  \bibinfo {author} {\bibfnamefont {L.}~\bibnamefont {Ranzani}}, \ and\
  \bibinfo {author} {\bibfnamefont {A.}~\bibnamefont {Kamal}},\ }\href
  {\doibase 10.1103/PhysRevResearch.2.023370} {\bibfield  {journal} {\bibinfo
  {journal} {Phys. Rev. Res.}\ }\textbf {\bibinfo {volume} {2}},\ \bibinfo
  {pages} {023370} (\bibinfo {year} {2020})}\BibitemShut {NoStop}%
\bibitem [{\citenamefont {Barreiro}\ \emph {et~al.}(2011)\citenamefont
  {Barreiro}, \citenamefont {M{\"u}ller}, \citenamefont {Schindler},
  \citenamefont {Nigg}, \citenamefont {Monz}, \citenamefont {Chwalla},
  \citenamefont {Hennrich}, \citenamefont {Roos}, \citenamefont {Zoller},\ and\
  \citenamefont {Blatt}}]{barreiro_open-system_2011}%
  \BibitemOpen
  \bibfield  {author} {\bibinfo {author} {\bibfnamefont {J.~T.}\ \bibnamefont
  {Barreiro}}, \bibinfo {author} {\bibfnamefont {M.}~\bibnamefont
  {M{\"u}ller}}, \bibinfo {author} {\bibfnamefont {P.}~\bibnamefont
  {Schindler}}, \bibinfo {author} {\bibfnamefont {D.}~\bibnamefont {Nigg}},
  \bibinfo {author} {\bibfnamefont {T.}~\bibnamefont {Monz}}, \bibinfo {author}
  {\bibfnamefont {M.}~\bibnamefont {Chwalla}}, \bibinfo {author} {\bibfnamefont
  {M.}~\bibnamefont {Hennrich}}, \bibinfo {author} {\bibfnamefont {C.~F.}\
  \bibnamefont {Roos}}, \bibinfo {author} {\bibfnamefont {P.}~\bibnamefont
  {Zoller}}, \ and\ \bibinfo {author} {\bibfnamefont {R.}~\bibnamefont
  {Blatt}},\ }\href {\doibase 10.1038/nature09801} {\bibfield  {journal}
  {\bibinfo  {journal} {Nature}\ }\textbf {\bibinfo {volume} {470}},\ \bibinfo
  {pages} {486} (\bibinfo {year} {2011})}\BibitemShut {NoStop}%
\bibitem [{\citenamefont {Reiter}\ \emph {et~al.}(2017)\citenamefont {Reiter},
  \citenamefont {S{\o}rensen}, \citenamefont {Zoller},\ and\ \citenamefont
  {Muschik}}]{reiter_autonomous_2017}%
  \BibitemOpen
  \bibfield  {author} {\bibinfo {author} {\bibfnamefont {F.}~\bibnamefont
  {Reiter}}, \bibinfo {author} {\bibfnamefont {A.~S.}\ \bibnamefont
  {S{\o}rensen}}, \bibinfo {author} {\bibfnamefont {P.}~\bibnamefont {Zoller}},
  \ and\ \bibinfo {author} {\bibfnamefont {C.~A.}\ \bibnamefont {Muschik}},\
  }\href {\doibase 10.1038/s41467-017-01895-5} {\bibfield  {journal} {\bibinfo
  {journal} {Nat. Commun.}\ }\textbf {\bibinfo {volume} {8}},\ \bibinfo {pages}
  {1822} (\bibinfo {year} {2017})}\BibitemShut {NoStop}%
\bibitem [{\citenamefont {Wang}\ and\ \citenamefont
  {Gertler}(2019)}]{wang_autonomous_2019}%
  \BibitemOpen
  \bibfield  {author} {\bibinfo {author} {\bibfnamefont {C.}~\bibnamefont
  {Wang}}\ and\ \bibinfo {author} {\bibfnamefont {J.~M.}\ \bibnamefont
  {Gertler}},\ }\href {\doibase 10.1103/PhysRevResearch.1.033198} {\bibfield
  {journal} {\bibinfo  {journal} {Phys. Rev. Res.}\ }\textbf {\bibinfo {volume}
  {1}},\ \bibinfo {pages} {033198} (\bibinfo {year} {2019})}\BibitemShut
  {NoStop}%
\bibitem [{\citenamefont {Verstraete}\ \emph {et~al.}(2009)\citenamefont
  {Verstraete}, \citenamefont {Wolf},\ and\ \citenamefont
  {Ignacio~Cirac}}]{verstraete_quantum_2009}%
  \BibitemOpen
  \bibfield  {author} {\bibinfo {author} {\bibfnamefont {F.}~\bibnamefont
  {Verstraete}}, \bibinfo {author} {\bibfnamefont {M.~M.}\ \bibnamefont
  {Wolf}}, \ and\ \bibinfo {author} {\bibfnamefont {J.}~\bibnamefont
  {Ignacio~Cirac}},\ }\href {\doibase 10.1038/nphys1342} {\bibfield  {journal}
  {\bibinfo  {journal} {Nat. Phys.}\ }\textbf {\bibinfo {volume} {5}},\
  \bibinfo {pages} {633} (\bibinfo {year} {2009})}\BibitemShut {NoStop}%
\bibitem [{\citenamefont {Zapusek}\ \emph {et~al.}(2023)\citenamefont
  {Zapusek}, \citenamefont {Javadi},\ and\ \citenamefont
  {Reiter}}]{zapusek_nonunitary_2023}%
  \BibitemOpen
  \bibfield  {author} {\bibinfo {author} {\bibfnamefont {E.}~\bibnamefont
  {Zapusek}}, \bibinfo {author} {\bibfnamefont {A.}~\bibnamefont {Javadi}}, \
  and\ \bibinfo {author} {\bibfnamefont {F.}~\bibnamefont {Reiter}},\ }\href
  {\doibase 10.1088/2058-9565/ac98dd} {\bibfield  {journal} {\bibinfo
  {journal} {Quantum Sci. Technol.}\ }\textbf {\bibinfo {volume} {8}},\
  \bibinfo {pages} {015001} (\bibinfo {year} {2023})}\BibitemShut {NoStop}%
\bibitem [{\citenamefont {Brandl}\ \emph {et~al.}(2016)\citenamefont {Brandl},
  \citenamefont {{van Mourik}}, \citenamefont {Postler}, \citenamefont {Nolf},
  \citenamefont {Lakhmanskiy}, \citenamefont {Paiva}, \citenamefont
  {M{\"o}ller}, \citenamefont {Daniilidis}, \citenamefont {H{\"a}ffner},
  \citenamefont {Kaushal} \emph {et~al.}}]{brandl_cryogenic_2016}%
  \BibitemOpen
  \bibfield  {author} {\bibinfo {author} {\bibfnamefont {M.~F.}\ \bibnamefont
  {Brandl}}, \bibinfo {author} {\bibfnamefont {M.~W.}\ \bibnamefont {{van
  Mourik}}}, \bibinfo {author} {\bibfnamefont {L.}~\bibnamefont {Postler}},
  \bibinfo {author} {\bibfnamefont {A.}~\bibnamefont {Nolf}}, \bibinfo {author}
  {\bibfnamefont {K.}~\bibnamefont {Lakhmanskiy}}, \bibinfo {author}
  {\bibfnamefont {R.~R.}\ \bibnamefont {Paiva}}, \bibinfo {author}
  {\bibfnamefont {S.}~\bibnamefont {M{\"o}ller}}, \bibinfo {author}
  {\bibfnamefont {N.}~\bibnamefont {Daniilidis}}, \bibinfo {author}
  {\bibfnamefont {H.}~\bibnamefont {H{\"a}ffner}}, \bibinfo {author}
  {\bibfnamefont {V.}~\bibnamefont {Kaushal}},  \emph {et~al.},\ }\href
  {\doibase 10.1063/1.4966970} {\bibfield  {journal} {\bibinfo  {journal} {Rev.
  Sci. Instrum.}\ }\textbf {\bibinfo {volume} {87}},\ \bibinfo {pages} {113103}
  (\bibinfo {year} {2016})}\BibitemShut {NoStop}%
\bibitem [{Note1()}]{Note1}%
  \BibitemOpen
  \bibinfo {note} {The equations describing the dynamics of the system are
  given in the Supplemental material, Section A}\BibitemShut {NoStop}%
\bibitem [{\citenamefont {Jost}\ \emph {et~al.}(2009)\citenamefont {Jost},
  \citenamefont {Home}, \citenamefont {Amini}, \citenamefont {Hanneke},
  \citenamefont {Ozeri}, \citenamefont {Langer}, \citenamefont {Bollinger},
  \citenamefont {Leibfried},\ and\ \citenamefont
  {Wineland}}]{jost_entangled_2009}%
  \BibitemOpen
  \bibfield  {author} {\bibinfo {author} {\bibfnamefont {J.~D.}\ \bibnamefont
  {Jost}}, \bibinfo {author} {\bibfnamefont {J.~P.}\ \bibnamefont {Home}},
  \bibinfo {author} {\bibfnamefont {J.~M.}\ \bibnamefont {Amini}}, \bibinfo
  {author} {\bibfnamefont {D.}~\bibnamefont {Hanneke}}, \bibinfo {author}
  {\bibfnamefont {R.}~\bibnamefont {Ozeri}}, \bibinfo {author} {\bibfnamefont
  {C.}~\bibnamefont {Langer}}, \bibinfo {author} {\bibfnamefont {J.~J.}\
  \bibnamefont {Bollinger}}, \bibinfo {author} {\bibfnamefont {D.}~\bibnamefont
  {Leibfried}}, \ and\ \bibinfo {author} {\bibfnamefont {D.~J.}\ \bibnamefont
  {Wineland}},\ }\href {\doibase 10.1038/nature08006} {\bibfield  {journal}
  {\bibinfo  {journal} {Nature}\ }\textbf {\bibinfo {volume} {459}},\ \bibinfo
  {pages} {683} (\bibinfo {year} {2009})}\BibitemShut {NoStop}%
\bibitem [{\citenamefont {Home}(2013)}]{home_quantum_2013}%
  \BibitemOpen
  \bibfield  {author} {\bibinfo {author} {\bibfnamefont {J.~P.}\ \bibnamefont
  {Home}},\ }in\ \href {\doibase 10.1016/B978-0-12-408090-4.00004-9} {\emph
  {\bibinfo {booktitle} {Advances {{In Atomic}}, {{Molecular}}, and {{Optical
  Physics}}}}},\ Vol.~\bibinfo {volume} {62}\ (\bibinfo  {publisher}
  {{Elsevier}},\ \bibinfo {year} {2013})\ pp.\ \bibinfo {pages}
  {231--277}\BibitemShut {NoStop}%
\bibitem [{\citenamefont {Brownnutt}\ \emph {et~al.}(2015)\citenamefont
  {Brownnutt}, \citenamefont {Kumph}, \citenamefont {Rabl},\ and\ \citenamefont
  {Blatt}}]{Brownnutt2015}%
  \BibitemOpen
  \bibfield  {author} {\bibinfo {author} {\bibfnamefont {M.}~\bibnamefont
  {Brownnutt}}, \bibinfo {author} {\bibfnamefont {M.}~\bibnamefont {Kumph}},
  \bibinfo {author} {\bibfnamefont {P.}~\bibnamefont {Rabl}}, \ and\ \bibinfo
  {author} {\bibfnamefont {R.}~\bibnamefont {Blatt}},\ }\href {\doibase
  10.1103/RevModPhys.87.1419} {\bibfield  {journal} {\bibinfo  {journal} {Rev.
  Mod. Phys.}\ }\textbf {\bibinfo {volume} {87}},\ \bibinfo {pages} {1419}
  (\bibinfo {year} {2015})}\BibitemShut {NoStop}%
\bibitem [{Note2()}]{Note2}%
  \BibitemOpen
  \bibinfo {note} {An analytic expression for the error is calculated in the
  Supplemental material, Section B}\BibitemShut {NoStop}%
\bibitem [{\citenamefont {Boldin}\ \emph {et~al.}(2018)\citenamefont {Boldin},
  \citenamefont {Kraft},\ and\ \citenamefont
  {Wunderlich}}]{boldin_measuring_2018}%
  \BibitemOpen
  \bibfield  {author} {\bibinfo {author} {\bibfnamefont {I.~A.}\ \bibnamefont
  {Boldin}}, \bibinfo {author} {\bibfnamefont {A.}~\bibnamefont {Kraft}}, \
  and\ \bibinfo {author} {\bibfnamefont {C.}~\bibnamefont {Wunderlich}},\
  }\href {\doibase 10.1103/PhysRevLett.120.023201} {\bibfield  {journal}
  {\bibinfo  {journal} {Phys. Rev. Lett.}\ }\textbf {\bibinfo {volume} {120}},\
  \bibinfo {pages} {023201} (\bibinfo {year} {2018})}\BibitemShut {NoStop}%
\bibitem [{\citenamefont {Home}\ \emph {et~al.}(2011)\citenamefont {Home},
  \citenamefont {Hanneke}, \citenamefont {Jost}, \citenamefont {Leibfried},\
  and\ \citenamefont {Wineland}}]{home_normal_2011}%
  \BibitemOpen
  \bibfield  {author} {\bibinfo {author} {\bibfnamefont {J.~P.}\ \bibnamefont
  {Home}}, \bibinfo {author} {\bibfnamefont {D.}~\bibnamefont {Hanneke}},
  \bibinfo {author} {\bibfnamefont {J.~D.}\ \bibnamefont {Jost}}, \bibinfo
  {author} {\bibfnamefont {D.}~\bibnamefont {Leibfried}}, \ and\ \bibinfo
  {author} {\bibfnamefont {D.~J.}\ \bibnamefont {Wineland}},\ }\href {\doibase
  10.1088/1367-2630/13/7/073026} {\bibfield  {journal} {\bibinfo  {journal}
  {New J. Phys.}\ }\textbf {\bibinfo {volume} {13}},\ \bibinfo {pages} {073026}
  (\bibinfo {year} {2011})}\BibitemShut {NoStop}%
\bibitem [{\citenamefont {Joshi}\ \emph {et~al.}(2020)\citenamefont {Joshi},
  \citenamefont {Fabre}, \citenamefont {Maier}, \citenamefont {Brydges},
  \citenamefont {Kiesenhofer}, \citenamefont {Hainzer}, \citenamefont {Blatt},\
  and\ \citenamefont {Roos}}]{Joshi2020}%
  \BibitemOpen
  \bibfield  {author} {\bibinfo {author} {\bibfnamefont {M.~K.}\ \bibnamefont
  {Joshi}}, \bibinfo {author} {\bibfnamefont {A.}~\bibnamefont {Fabre}},
  \bibinfo {author} {\bibfnamefont {C.}~\bibnamefont {Maier}}, \bibinfo
  {author} {\bibfnamefont {T.}~\bibnamefont {Brydges}}, \bibinfo {author}
  {\bibfnamefont {D.}~\bibnamefont {Kiesenhofer}}, \bibinfo {author}
  {\bibfnamefont {H.}~\bibnamefont {Hainzer}}, \bibinfo {author} {\bibfnamefont
  {R.}~\bibnamefont {Blatt}}, \ and\ \bibinfo {author} {\bibfnamefont {C.~F.}\
  \bibnamefont {Roos}},\ }\href {\doibase 10.1088/1367-2630/abb912} {\bibfield
  {journal} {\bibinfo  {journal} {New Journal of Physics}\ }\textbf {\bibinfo
  {volume} {22}},\ \bibinfo {pages} {103013} (\bibinfo {year}
  {2020})}\BibitemShut {NoStop}%
\bibitem [{\citenamefont {Morigi}\ \emph {et~al.}(2000)\citenamefont {Morigi},
  \citenamefont {Eschner},\ and\ \citenamefont {Keitel}}]{Morigi2000}%
  \BibitemOpen
  \bibfield  {author} {\bibinfo {author} {\bibfnamefont {G.}~\bibnamefont
  {Morigi}}, \bibinfo {author} {\bibfnamefont {J.}~\bibnamefont {Eschner}}, \
  and\ \bibinfo {author} {\bibfnamefont {C.~H.}\ \bibnamefont {Keitel}},\
  }\href {\doibase 10.1103/PhysRevLett.85.4458} {\bibfield  {journal} {\bibinfo
   {journal} {Phys. Rev. Lett.}\ }\textbf {\bibinfo {volume} {85}},\ \bibinfo
  {pages} {4458} (\bibinfo {year} {2000})}\BibitemShut {NoStop}%
\end{thebibliography}%
\clearpage



\def\theequation{S\arabic{equation}}
\def\thefigure{S\arabic{figure}}
\def\thetable{S\arabic{table}}
\setcounter{figure}{0}

\begin{center} \textbf{\large{Supplemental material for \\
\emph{Experimental realization of nonunitary multi-qubit operations}}}
\end{center}

\section{System}
\label{sec:AppendixSystem}
The dynamics of our system are well described by a master equation of Lindblad form. The Hamiltonian term $\hat{H}$ describes the coherent interaction while the dissipative processes are described by the Lindblad jump operators $\hat{L}_k$:
\begin{equation}\label{eq:lind}
    \Dot{\rho} = - i [\hat{H},\rho] + \sum_k \hat{L}_k \rho \hat{L}^\dag_k - \frac{1}{2}(\hat{L}^\dag_k \hat{L}_k\rho + \rho \hat{L}^\dag_k \hat{L}_k).
\end{equation}
 We consider three subsystems: The electronic state of the first ion $k$, that of the second ion $l$, and a common motional mode $m$ occupied with $n$ phonons, we use the notation $\bracket{k,n}_1 \otimes \bracket{l,n}_2=\bracket{kl}\bracket{n}_m$. When referring to single-qubit couplings we write $\ketbraind{a}{j}{b}$ where $j$ indicates the addressed ion. The system Hamiltonian can be written as:
\begin{equation}
    \hat{H} = \delta \hat{a}^\dag \hat{a}+\hat{H}_{f,1}+\hat{H}_{f,2}+\hat{V}_1.
\end{equation}
Here, $\hat{a}^{(\dagger)}$ are the annihilation (creation) operators of the oscillator mode. The subscripts of the operators denote the qubit the respective coupling acts on.

We apply a carrier drive, referred to as the probe, on the $\ket{0}\leftrightarrow\ket{f}$ transition
\begin{equation}
    \hat{V}_1= \frac{\Omega_f}{2} (\ketbraind{f}{1}{0}+\ketbraind{0}{1}{f}). 
\end{equation}

On the $\ket{f}\leftrightarrow\ket{1}$ transition we apply a sideband drive
\begin{equation}
    \hat{H}_{f,j}= \frac{\Omega_{\m{SB}}}{2}(\hat{a}\ketbraind{f}{j}{1} + \hat{a}^\dag\ketbraind{1}{j}{f}) + \Delta \ketbraind{f}{j}{f}.
\end{equation}
The dissipative contribution used for both OR and NOR is sympathetic cooling. It allows the excited state of the first qubit to decay into the state $\ket{1}$, represented by the following Lindblad operator:
\begin{equation}
    \hat{L}_{\Gamma_f}=\sqrt{\Gamma_f}\hat{a}.
\end{equation}
The NOR gate additionally uses decay over an auxiliary level:
\begin{equation}
    \hat{L}_{\Gamma_e}=\sqrt{\Gamma_e}\ketbraind{0}{j}{e}.
\end{equation}
This coupling is present on both qubits.

\section{Analytic error analysis}
\label{sec:AppendixAnalytic}
In principle the logic gates presented here can operate at arbitrary low error. However, due to experimental imperfections and limited coupling strength we observe errors.
We analyse the error processes for the OR gate and derive an analytic approximation of the error to better understand these imperfections. 

The experiment operates in separate steps for excitation and decay. For this analysis we will assume that the decay by sympathetic cooling operates perfectly and that all population that was excited by the Rabi pulse decays to the target state.

The main source of error is heating. It competes with sympathetic cooling to raise the occupation number of the harmonic oscillator. After state preparation the harmonic oscillator is in a thermal state with a mean phonon number $\Bar{n}=0.14$. 
For this low occupation number the probability to be in Fock state $\ket{0}_m$ is $P_0 =1/(1+\Bar{n})$ and that for Fock state $\ket{1}_m$ is $P_1 \approx 1 -P_0$.

In the first step population from $\ket{01}\ket{0}_m$ is transferred to the dressed state $\ket{\psi_-} = (\ket{f1}\ket{0}_m+\ket{1f}\ket{0}_m-\sqrt{2}\ket{11}\ket{1}_m)/2$ by a resonant $\pi/2$ pulse. This dressed state is addressed by the effective Rabi frequency $\Omega_f/2$. Therefore the pulse takes $t_{\pi} = 2\pi/\Omega_f$.

\begin{figure*}
    \centering
    \includegraphics{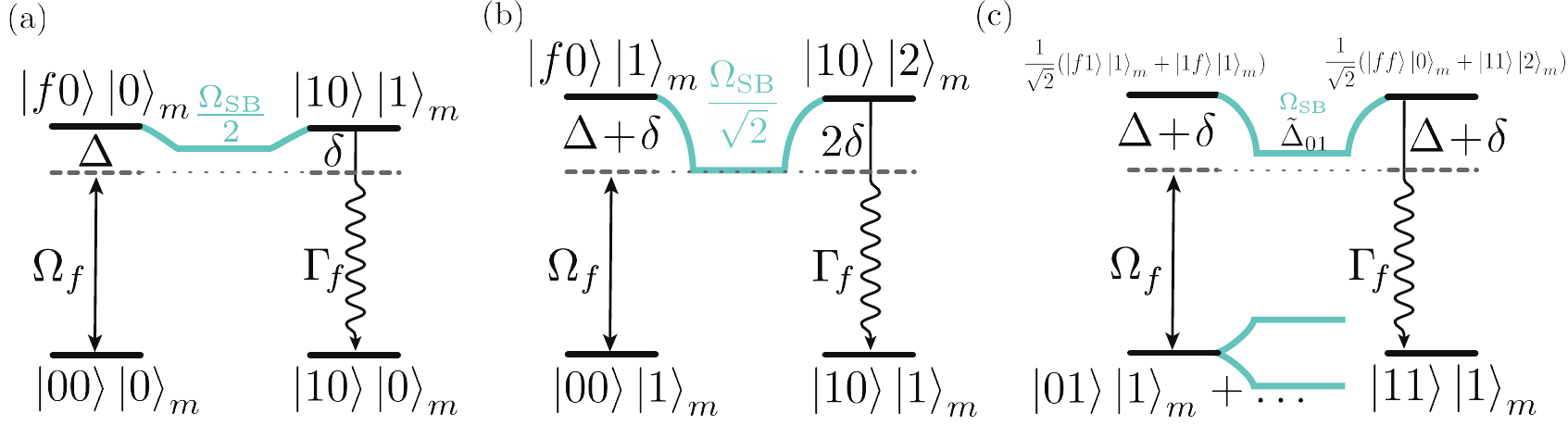}
    \caption{Level schemes of the error processes. (a) The error process intrinsic to the scheme. Even in absence of any harmonic oscillator excitations some population from $\ket{00}\ket{0}_m$ is off-resonantly excited. By increasing the sideband frequency $\Omega_\text{SB}$ relative to the other couplings this error could be suppressed. (b) If the harmonic oscillator is in state $n=1$ the energies of the excited states addressed from $\ket{00}\ket{1}_m$ are shifted by $\delta$ and the coupling between them is enhanced. This shifts the probe into resonance. (c) The desired process in presence of an harmonic oscillator excitation. The additional excitation enables coupling to the double excited state $\ket{ff}\ket{0}_m$ and adjusts the frequencies to $\Delta+\delta$. Furthermore, $\ket{01}\ket{1}_m$ forms dressed states with $\ket{0f}\ket{0}_m$. The lowest frequency dressed state resides at $\tilde{\Delta}_{01}$.}
    \label{fig:Error_levels}
\end{figure*}

The same pulse that resonantly excites $\ket{01}\ket{0}_m$ also addresses state $\ket{00}\ket{0}_m$, off-resonantly driving it to $\ket{\phi_-} = (\ket{f0}\ket{0}_m-\ket{10}\ket{1}_m)/\sqrt{2}$. It is detuned by $\Delta_d = \Omega_\text{SB}(\sqrt{2}-1)/2$ and the Rabi frequency it is addressed with is $\Omega_d = \Omega_f/\sqrt{2}$. Using these values we can calculate the excitation probability just as we would for a simple detuned Rabi pulse:
\begin{equation}\label{eq:excitation_Rabi}
   P_e^{00}(n=0) = \frac{\Omega_d^2}{\Delta_d^2+\Omega_d^2} \sin{\left(\frac{1}{2}\sqrt{\Delta_d^2+\Omega_d^2} t_{\pi/2}\right)} \approx 17\%.
\end{equation}
This process is shown in Fig.~\ref{fig:Error_levels}(a).
If the system initially is in $n=1$ the dressed states are resonant and addressed by Rabi frequency $\Omega_d = \Omega_f/\sqrt{2}$ (see Fig. \ref{fig:Error_levels}(b)). The total error probability we observe is:
\begin{equation}
    P_e^{00} = P_0 P_e^{00}(n=0) + P_1 P_e^{00}(n=1) \approx 23\%.
\end{equation}
In the experiment we measure an error of $P_e^{00} = 14\%$. 

\begin{table}[]
    \centering
    \begin{tabularx}{\columnwidth}{|l|X|X|X|}
    \hline
    Initial state &  Measured & Numeric  &Analytic \\
    \hline
     $\ket{00}$& 86 & 79 & 79\\
    \hline
       $\ket{01}$& 84 & 86 & 82 \\
    
     \hline
    
    \end{tabularx}
    \caption{Measured fidelity compared to analytic approximations for input states $\ket{00}$ and $\ket{01}$. The fidelity of 0.98 of the initialization explains part of the difference between analytics and experiment. Initial states $\ket{10}$ and $\ket{11}$ are not shown as their fidelity is close to unity.}
    \label{tab:Comparison}
\end{table}

When the system is in initial state $\ket{01}\ket{1}_m$ it is driven to an excited subspace consisting of states $\ket{f1}\ket{1}_m, \ket{11}\ket{2}_m, \ket{1f}\ket{1}_m$ and $\ket{ff}\ket{0}_m$ that couple to each other (cf. Fig. \ref{fig:Error_levels} (c)). Furthermore, the sideband drive couples state $\ket{01}\ket{1}_m$ to $\ket{0f}\ket{0}_m$ forming dressed states $\ket{\psi_\pm} = (\ket{01}\ket{1}_m \pm \ket{0f}\ket{0}_m)/\sqrt{2}$. 
State $\ket{\psi_-}$ is close to resonance, driven with $\tilde{\Delta }_{01-}= (1+\sqrt{2}-\sqrt{6})\Omega_\text{SB}/2$ and $\Omega_d = \Omega_f (1/\sqrt{8} + 1/\sqrt{12})$. Using Eq. \eqref{eq:excitation_Rabi} one can find the error probability for this initial state to be about $24 \%$. State $\ket{\psi_+}$ sits at detuning  $\tilde{\Delta }_{01+}= (-1+\sqrt{2})\Omega_\text{SB}/2$ with respect to the excited state manifold and is addressed by Rabi frequency $\Omega_d = \Omega_f/2$. Therefore it has an error probability of $89\%$. Combining both error probabilities with the probability to start in an excited state of the oscillator, the total error probability for initial state $\ket{01}$ evaluates to
\begin{equation}
    P_e^{01} = P_e^{01}(n=1)(1- P_0) \approx 7\%.
\end{equation}
This value is lower than the experimental error of  $P_e^{01} = 16\%$. The discrepancy is explained as follows:

We have independently measured that experimental imperfections lead to a Rabi cycle decay that can be modeled by a Gaussian statistical spread in Rabi frequency with a standard deviation of $4 \%$. Including this spread in our gate error models gives the results shown in Tab. \ref{tab:Comparison}.

For both initial states $\ket{00}$ and $\ket{01}$ we overestimate the population that is transferred to the excited states. Here we only took the initial occupation of the oscillator into account while in the experimental setting heating also occurs during the $\pi/2$ pulse. Additionally, the state initialization for states other than $\ket{00}$ has a fidelity of 98\%. Another source of error could be miscalibration of pulse timing. Furthermore, we do not consider errors during the decay step. These effects combined can explain the difference between the experiment and numerical as well as analytical models. Such sources of error are technical in nature, and therefore not a fundamental limitation. This provides a clear path how to improve the gates' performance.

\end{document}